\documentclass[useAMS,fleqn,usenatbib]{mnras}

\usepackage{amsmath}	
\usepackage{amssymb}	

\usepackage{txfonts}

\usepackage[T1]{fontenc}
\usepackage{ae,aecompl}

\usepackage{graphicx}	

\usepackage{multicol}        
\usepackage{bm}		
\usepackage{color}
\usepackage{rotating}

\usepackage{xcolor}

\usepackage{siunitx}

\DeclareSIUnit \h {\ensuremath{\mathit{h}}}
\DeclareSIUnit \parsec {pc}

\def\be{\begin{equation}}
\def\ee{\end{equation}}
\def\ba#1\ea{\begin{align}#1\end{align}}

\newcommand{\refeq}[1]{Eq.~(\ref{eq:#1})}          
\newcommand{\refeqs}[2]{Eqs.~(\ref{eq:#1})--(\ref{eq:#2})}

\def\abg{a}

\def\Hbgn{H_\mathrm{0}}

\def\comment#1{}

\newcommand{\MPA}{Max Planck Institute for Astrophysics}

\newcommand{\Monofonic}{\textsc{monofonic-MUSIC}2\,}
\newcommand{\Gadget}{\textsc{gadget}}
\newcommand{\gadget}{\textsc{gadget}4\,}

\newcommand{\hlm}[1]{#1}
\newcommand{\hlt}[1]{#1}
\newcommand{\hlcom}[1]{}

\newcommand{\myvec}[1]{{ \bm{#1} }}

\def\psid{\dot{\Psi}}
\def\psidd{\ddot{\Psi}}

\makeatletter
\newcommand{\printfnsymbol}[1]{%
  \@fnsymbol{#1}
}
\makeatother

\title[Tidal Response]{Measuring the Tidal Response of 
Structure Formation:\\
Anisotropic Separate Universe Simulations using TreePM}
\author[J. St\"ucker et al.]{Jens St\"ucker$^{1}$\thanks{These authors contributed equally.}\thanks{E-mail: jstuecker@dipc.org}, 
Andreas. S. Schmidt$^{2}$\printfnsymbol{1},
Simon D. M. White$^{2}$, 
\newauthor
Fabian Schmidt$^{2}$ and Oliver Hahn$^{3}$
\\
$^{1}$Donostia International Physics Centre (DIPC), Paseo Manuel de Lardizabal 4, 20018 Donostia-San Sebastian, Spain.\\
$^{2}$\MPA , Karl-Schwarzschild-Str. 1, 85741 Garching, Germany\\
$^{3}$Laboratoire Lagrange, Universit\'e C\^ote d'Azur, Observatoire de la C\^ote d'Azur, CNRS,
      Blvd de l'Observatoire,\\\hskip0.15in CS 34229, 06304 Nice cedex 4, France.\\
}
\date{Accepted XXX. Received YYY; in original form ZZZ}

\pubyear{2019}

\begin{document}
\label{firstpage}
\pagerange{\pageref{firstpage}--\pageref{lastpage}}
\maketitle

\begin{abstract}
  We present anisotropic ``separate universe'' simulations which modify the N-body code \gadget in order to represent a large-scale tidal field through an anisotropic expansion factor.
These simulations are used to measure the linear, quasi-linear and nonlinear response of the matter power spectrum to a spatially uniform trace-free tidal field up to wavenumber $k = \hlt{\SI{7}{\h\per\mega\parsec}}$. Together with the response to a large-scale overdensity measured in previous work, this completely describes the nonlinear matter bispectrum in the squeezed limit.
We find that the response amplitude does not approach zero on small scales in physical
coordinates, but rather a constant value \hlt{at $z=0$}, $R_K\approx 0.5$ for $k \geq \SI{3}{\h\per\mega\parsec}$ \hlt{up to the scale where we consider our simulations reliable, $k \leq \SI{7}{\h\per\mega\parsec}$}. This shows that even the inner regions of haloes are affected by the large-scale tidal field.
We also measure directly the alignment of halo shapes with the tidal field, finding a clear signal which increases with halo mass. 
\end{abstract}

\begin{keywords}
methods: numerical - cosmology: large-scale structure of Universe.
\end{keywords}



\section{Introduction}

Large-scale tidal fields influence the growth of structure in a characteristic, anisotropic way. This applies to both quasilinear and fully nonlinear scales. On quasilinear to nonlinear scales, this effect can be observed as an anisotropy in the matter power spectrum which aligns with the large-scale tidal field,
an effect described via the so-called \emph{power spectrum response}. 
\citep{r1,r2,r4,r5,wagner/etal:2014,2017JCAP...06..053B}. This coupling of the small-scale matter power spectrum to
large-scale perturbations leads to an important contribution to the covariance
of weak gravitational lensing statistics, which probe the projected
matter distribution \citep{takada/hu:2013,li/hu/takada:2014,responses2,Barreira/Krause/Schmidt:2017}\hlt{, as well as to galaxy clustering statistics} \hlt{\citep{Akitsu:2016leq,Li:2017qgh,Chiang:2018mau}}. 
On even smaller scales, the shapes of galaxies and galaxy clusters, as well as of the haloes that host them, align with large-scale tides \citep[][see \citet{alignment_review} for a review]{Catelan01,Heavens00,Hirata04,hahn_2007,Schafer17}.
This effect is commonly known as \emph{intrinsic alignment}, and is an important contaminant when attempting to measure weak gravitational lensing through galaxy shape correlations. Thus, quantifying the alignment strength of haloes is an important ingredient in modeling galaxy shape statistics.

Beyond being important ingredients to be modeled when interpreting weak lensing observables, the impact of tidal fields can also be used as a cosmological probe: the small-scale power spectrum can be used to reconstruct large-scale perturbations \citep{pen/etal:2012,zhu/etal:2016,li/dodelson/croft:2020}. Similarly, intrinsic alignments themselves contain a wealth of cosmological information \citep{CD13,Chisari14,schmidt/chisari/dvorkin}. These techniques require knowledge of how the small-scale matter power spectrum and halo shapes, respectively, respond to large-scale tidal fields.

\begin{figure*}
    \centering
    \includegraphics[width=\textwidth]{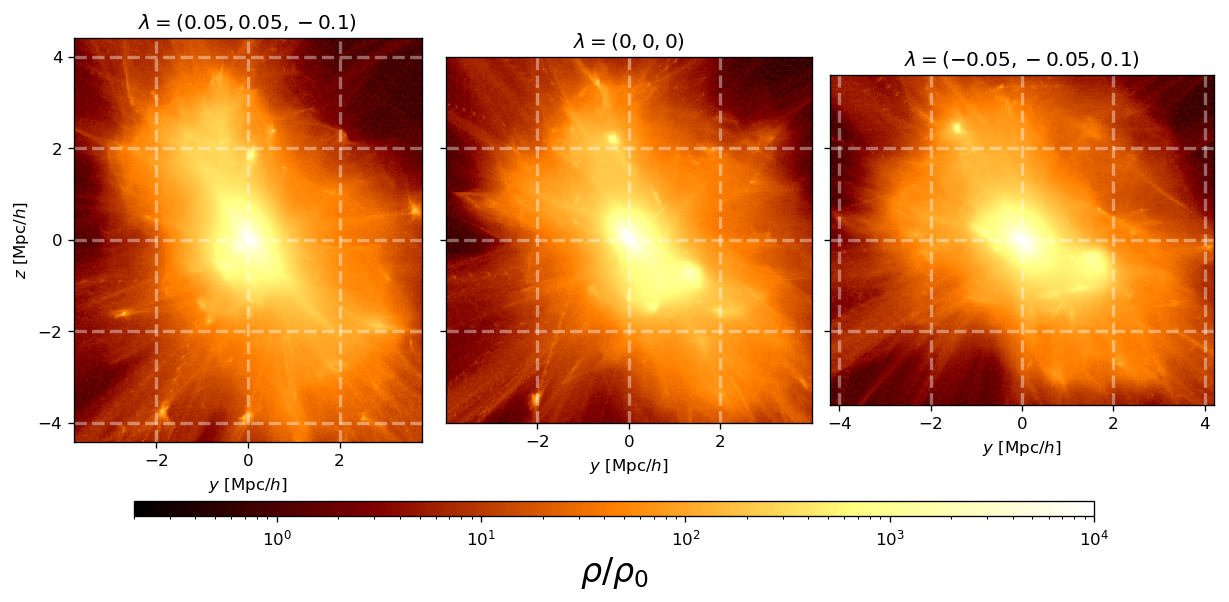}
    \caption{The same halo in three simulations using the same realisation of the initial conditions but different tidal fields. The haloes are shown in the physical frame. The coordinate grid (dashed lines) is in physical coordinates and is intended to facilitate comparison between different panels. In the case with a ''stretch'' along the z-axis  $\lambda_z = -0.1$ (left) the halo aligns more strongly with the $z$-axis (vertical axis in the plot) than in the case with opposite tidal field (right). The same effect can be seen in the results that are presented in section \ref{sec:Results}.
    The images have been created using trigonometric sheet resampling \citep{stuecker_2018}.}
    \label{fig:halo_vis}
\end{figure*}

In this paper, we introduce a tool to simulate precisely the impact of a large-scale tidal field on the growth of structure. Following \cite{Schmidt:2018}, we incorporate the tidal field in an N-body simulation via an anisotropic expansion factor. We extend the results of \cite{Schmidt:2018}, who showed results from a fixed-grid particle-mesh (PM) code, to solve for the full TreePM force. This allows us to follow the effect of the tidal field down to much smaller scales and to probe the interior of haloes. The main advantage of this simulation technique is that it allows us to impose a controlled, spatially uniform tidal field on the full N-body dynamics. By taking a numerical derivative with respect to this uniform tidal field, the tidal effects can be isolated very precisely, and cosmic variance is canceled to a large extent. This mirrors the very similar advantages of the separate-universe technique when simulating the effects of large-scale overdensities \citep{Frenk/White/Davies:1988,mcdonald:2003,sirko:2005,martino/sheth:2009,gnedin/kravtsov/rudd:2011,li/hu/takada:2014,wagner/etal:2014,baldauf/etal:2015,SU-TNG}. On the other hand, tidal effects in conventional N-body simulations are difficult to extract robustly, because tidal fields fluctuate and are present on all spatial scales. 

As an example, it is very difficult to determine to what extent the shape of a given halo in a standard N-body simulation is influenced by tidal fields on a given scale. The technique presented here however allows for a measurement of the tidal effect on a single halo, by performing N-body simulations using the same initial seeds both with and without an external tidal field. This is illustrated in Fig.~\ref{fig:halo_vis}, which shows slices through TreePM N-body simulations without a tidal field and with tidal fields of opposite sign. The tidal field here has a dimensionless amplitude (parametrized by $\lambda$) of $0.1$, corresponding to a universe with fractional differences in scale factor of roughly 10\% between the $z$ and $x,y$ axes. The effect on the shape of the halo in the centre of the figure is clearly visible, especially in its outer regions.

\hlcom{Added following paragraph + citation:} We note that the separate universe approach has similarities with the approach taken in the sCOLA scheme \citep{tassev2015scola, Leclercq_2020} which also separates the treatment of quasi-linear large scales and non-linear smaller scales. However, in the sCOLA scheme the goal is to make faster and bigger simulations whereas in our case the goal is to single out and understand the effect of the large-scale tidal field.

Following the two physics goals presented above, we show two applications in this paper: first, we measure the anisotropic response of the matter power spectrum to large-scale tidal fields, extending previous results up to \hlcom{$k = \SI{7}{\h\per\mega\parsec}$}.  This result can be immediately applied to the calculation of the covariance of weak lensing power spectra. 
Second, we show simple measurements of the alignment of halo shapes. Our very high signal-to-noise detection of this effect illustrates the power of our simulation technique for detailed halo alignment studies, as alluded to above and illustrated in Fig.~\ref{fig:halo_vis}: by measuring the alignment of haloes with respect to a fixed, external tidal field, the noise in the measurement is substantially reduced.

Sections \ref{sec:initialconditions}--\ref{sec:modGadget} present a short introduction to the large-scale tidal field equations, as well as the modifications we have made in the initial conditions generator and the N-body code \Gadget4 \hlt{\citep{gadget4paper}}. In section \ref{sec:Response} we show how to measure the response in simulations, and we demonstrate convergence of these measurements. The main results are presented in section \ref{sec:Results} -- these include the response measurement in the fully nonlinear case, and a brief study of halo alignment with the tidal field. Several appendices give technical details of our procedures.
\section{Lagrangian Perturbation Theory in the anisotropic frame - the generation of initial conditions} \label{sec:initialconditions}

To measure the response reliably at all times, we have to make sure that both the initial conditions and the simulation code account properly for evolution in the anisotropic frame. In this section we will discuss how Lagrangian Perturbation Theory (LPT) can be applied in the anisotropic frame and what modifications are needed to existing 2LPT codes to generate initial conditions in the anisotropic frame. We discuss the main steps in the next subsection, but relegate the full derivation to Appendix \ref{lpt:appendix}. \hlcom{Updated the following sentence:} These modifications have been implemented into the IC code \Monofonic \citep{MUSIC2monofonic,Michaux:2020}, which is publicly available\footnote{Available from \url{https://bitbucket.org/ohahn/monofonic}.}.


\subsection{Expansion of an Anisotropic Universe} \label{sec:background}

\cite{Schmidt:2018} describe how the evolution of a universe with a homogeneous large-scale tidal field can be modelled by an N-body simulation with periodic boundary conditions. A general tidal field is described by a symmetric 3x3 tensor, but we can rotate into principal axis coordinates where the tidal tensor is diagonal. In this frame we can describe the anisotropic expansion by three scale factors $a_x$, $a_y$ and $a_z$, which replace the single $a(t)$ of the standard isotropic model.\footnote{Note that this assumes the orientation of the external tidal field to be constant in time. Only its eigenvalues can change. In a perturbative expansion of the large-scale tidal field, This is the case up to including second order. Equivalently, this holds as long as the external large-scale mass distribution which generates the tidal field can be described as evolving according to 2LPT.}

For our purpose, it is useful to think of this anisotropic simulation as a small region of a larger isotropically expanding universe that grows according to a background scale factor $\abg$ (with today's epoch corresponding to $\abg = 1$). It is convenient to describe the scale evolution of the anisotropic region with respect to the isotropic background universe by considering expansion factor ratios $\alpha_i = a_i / \abg$.  The region simulated evolves differently from the isotropic background in such a way as to mimic the dynamical effects of a large-scale tidal field. The scale-factor ratios are defined to approach unity in the limit of early times $\alpha_i(\abg \rightarrow 0) \rightarrow 1$, but they then deviate significantly from unity at later times. These ratios follow modified versions of the Friedman equations. As explained in \citet{Schmidt:2018}  and \citet{stuecker_2018}, these are given by

%

\begin{align}
\dot\alpha_i &= \abg^{-2} \eta_i\,, \label{eqn:backgrounddrift}\\
\dot\eta_i &= - \frac32 \Omega_{\text{m},0} \Hbgn^2 \abg^{-1}(t) \alpha_i(t) \Lambda_i(t)\,,
\label{eq:alphaEOMi}
\end{align}
where the $\eta_i$ are convenient momentum variables and the $\Lambda_i$ are nonlinear versions of the eigenvalues of the deformation tensor which parametrise the external tidal field while also accounting for the fact that the nonlinear density of the simulation box is known exactly from the ratios $\alpha_i$. For our current work, we adopt a linear external tidal field, giving:
\begin{align}
    \Lambda_i(t) &= \hlm{\frac{1}{3} \left(\frac{1}{\alpha_1(t) \alpha_2(t) \alpha_3(t)} - 1 \right)} - D_1(t) \left( \lambda_i - \frac{1}{3} \sum_j \lambda_j \right)\,,\label{eqn:lineartidalfield}
\end{align}
where $D_1$ is the linear growth factor and $\lambda_i$ are the eigenvalues of the deformation tensor $d_{ij}$ of the large-scale perturbation \citep[compare appendix of][]{stuecker_2018}, and
\begin{align}
  d_{ij} &= \frac{\partial_i \partial_j}{\sum_k \partial_k^2} \delta_{\text{lin}}\,.
\end{align}
In our case we leave $\lambda_i$ as free parameters which determine the amplitude of the external perturbation. The deformation of the anisotropic universe is then given at early times by the Zeldovich approximation,
\begin{align}
    \alpha_i \approx 1 - D_1(t) \lambda_i\,, \label{eqn:approxevol}
\end{align}
which also defines the initial conditions for the numerical integration of the evolution equations. For the simulations in later sections, we numerically integrate the anisotropic background evolution using \eqref{eqn:backgrounddrift} - \eqref{eqn:lineartidalfield} to give an accurate treatment even in the nonlinear regime. However, for the purpose of this section, approximation \eqref{eqn:approxevol} is accurate enough to obtain Lagrangian Perturbation Theory initial conditions up to second order.

\subsection{The anisotropic comoving coordinate frame} \label{sec:comovingcoordinates}
We define anisotropic comoving coordinates $\myvec{x}$. Simulation particles which are not subject to any perturbations around the  local anisotropically expanding region stay at constant $\myvec{x}$. Therefore, these coordinates are the natural choice of ``comoving" coordinates in this context. However, to avoid confusion with the comoving coordinate frame normally defined in the literature (which is relative to the isotropic expansion of the larger scale background universe) we refer to the two different frames as ``anisotropic comoving" and ``isotropic comoving" frames in the following. The anisotropic comoving coordinates relate to physical coordinates $\myvec{r}_{ \text{phys}}$ and to isotropic comoving coordinates $\myvec{r}$ as 
\begin{align}
  r_{i, \text{phys}} &=  \abg r_{i} \,,\\
  r_{i} &= \alpha_i x_i \,.
\end{align}
Throughout the paper, we distinguish between anisotropic comoving coordinates $\myvec{x}$ and isotropic comoving coordinates $\myvec{r}$. All spatial derivatives are by default taken in anisotropic comoving coordinates:
\begin{align}
    \partial_i := \frac{\partial}{\partial x^i}\,.
\end{align}

\subsection{Equations of Motion}  \label{sec:equationsofmotion}

The equations of motion in the anisotropic comoving frame are obtained by
generalizing the well-known equations for a non-relativistic particle from
an isotropic to an anisotropic background:
\begin{align}
\dot{x}_i &= \frac{v_i}{a^2 \alpha_i^2} \label{eqn:motion_x}\,, \\
\dot{v}_i &= - \frac{\partial_i \phi} {a} \label{eqn:motion_v}\,, \\
\textrm{and}\,\,\sum_{i=1}^3 \alpha_i^{-2} \partial_i^2 \phi &=  \frac{4 \pi G \rho_0 \delta}{\alpha_1 \alpha_2 \alpha_3}\,, \label{eqn:poisson}
\end{align}
where $\myvec{v}$ are the canonical momenta associated with $\myvec{x}$, $\rho_0$ is the mean matter density \hlt{at $a=1$ in the isotropic frame} and $\delta(\myvec{x}) = \rho(\myvec{x})/\rho_0 - 1$ is the relative density contrast at a location $\myvec{x}$ \hlt{where $\rho(\myvec{x})$ is the comoving density} \citep{Schmidt:2018}.

This is a suitable form for solving the equations of motion numerically, as we will discuss in section \ref{sec:modGadget}. However, to apply perturbation theory it is more convenient to combine the two first-order equations of motion and the Poisson equation into a single second-order differential equation with one vector variable. This can be done by combining the time derivative of equation \eqref{eqn:motion_x} with \eqref{eqn:motion_v}, taking the divergence, and using the Poisson equation \eqref{eqn:poisson}:
\begin{align}
  \myvec{\nabla} \left( \ddot{\myvec{x}} + 2 H^*(a) \dot{\myvec{x}} \right) &= \hlm{-} \frac{4 \pi G \rho_0}{a^3 \alpha_1 \alpha_2 \alpha_3} \delta(\myvec{x}) \nonumber \\
                              &= \hlm{-} \frac{3 H^2 (a) \Omega_m(a)}{2 \alpha_1 \alpha_2 \alpha_3} \delta(\myvec{x}), \label{eqn:eqofmotionsingle}
\end{align}
where we have used the Friedman equation of the isotropic background universe, together with $\Omega_m(a) = \rho_m(a) / \rho_{\text{crit}}(a)$, and defined an anisotropic Hubble tensor which is diagonal in our case and has the components,
\begin{align}
  H^*_{ij}(a) = \delta_{ij} \left( \frac{\dot{a}}{a} + \frac{\dot{\alpha}_i}{\alpha_i} \right), \label{eqn:anisohubble}
\end{align}
where no summation over identical indices is implied, and $\delta_{ij}$ denotes the Kronecker delta.

\subsection{Lagrangian Perturbation Theory}

The Lagrangian Perturbation Theory solutions can be found by making the ansatz
\begin{align}
    \myvec{x}(\myvec{q},t) &= \myvec{q} + \myvec{\Psi}(\myvec{q}, t) ,\\
    \myvec{\Psi}(\myvec{q}, t) &\approx D_1(t) \myvec{\Psi}^{(1)}(\myvec{q}) + D_2(t) \myvec{\Psi}^{(2)}(\myvec{q})  + D_{2\lambda}(t) \myvec{\Psi}^{(2\lambda)}(\myvec{q}) + O(3).
\end{align}
where the solutions to $D_1$, $\myvec{\Psi}_1$, $D_2$ and $\myvec{\Psi}^{(2)}$ are the ones given by standard 2LPT,\footnote{Together with a correction accounting for the large-scale overdensity if $\sum \lambda_i \neq 0$} (i.e. without an external tidal field). In an Einstein-de Sitter universe, $D_2 = -(3/7)D_1^2$, a relation which is quite accurate even in $\Lambda$CDM. $D_{2\lambda}$ and $\myvec{\Psi}^{(2\lambda)}$ denote second order corrections which account for the anisotropic evolution and have a different time dependence than $D_2$. Notice that there is no $\myvec{\Psi}^{(1\lambda)}$ term, since we work in anisotropic comoving coordinates where the linear-order effect of the tidal field is already incorporated. We find that the solutions for the anisotropic corrections at second order are, 
\begin{align}
    D_{2\lambda} &= D_1^2 + D_2 ,\\
    \myvec{\nabla} \hlm{{\cdot}} \myvec{\Psi}^{(2\lambda)} &= \sum_i \lambda_i \partial_i \hlm{\Psi}^{(1)}_i,
\end{align}
as derived in Appendix \ref{lpt:appendix}. These solutions are correct for any $\Lambda$CDM universe and are consistent with previous predictions of the effects of a large-scale tidal field, as shown in Appendix \ref{app:response}.
\subsection{Generation of Initial Conditions}
To make minimal changes to existing initial condition generators, we additionally make the approximation
\begin{align}
    D_{2\lambda} &\approx - \frac{4}{3} D_2
\end{align}
which is exact in an Einstein de Sitter universe, and still has only an order 1\% error at $a=1$ for a universe similar to ours $\Omega_m(a=1) \sim 0.3$ as we show in Appendix \ref{lpt:appendix2lptapprox}. At early times, where initial conditions are typically generated, this error is completely negligible. By using this approximation, the only change with respect to standard 2LPT is a modification to the Poisson equation of the second order displacement potential to
\begin{align}
    \sum_i \partial_i^2 \Phi^{(2)} \rightarrow \sum_i \partial_i^2  \Phi^{(2)} \hlm{+} \sum_i \partial_i^2 \Phi^{(1)} \left(\sum_j \lambda_j \right)  \hlm{+} \frac{4}{3} \sum_i \lambda_i \partial_i^2 \Phi^{(1)}\,,
\end{align}
where the potentials generate the displacement fields via \hlt{${\Psi}^{(1)}_i = - \partial_i\Phi^{(1)}$ and $\hlm{{\Psi}^{(2)}_i} = \partial_i\Phi^{(2)}$}. 
In this modification we have also included an isotropic correction to $\Phi^{(2)}$ for cases with large-scale overdensities $\sum \lambda_i \neq 0$ so that this equation remains valid for arbitrary combinations of $\lambda_i$ (compare Appendix \ref{lpt:appendix}). We implemented this change and the correct transformation of the velocity variables
\begin{align}
    v_i &=a^2 \alpha_i^2  \dot{x}_i \,.
\end{align}
into \hlt{\Monofonic}. We use this for generating initial conditions for all simulations presented in this paper.

\section{Modified \textsc{gadget}4}
\label{sec:modGadget}
Here we describe the changes that are necessary to a TreePM N-body code to perform simulations in the anisotropic frame. The simulation volume is chosen to be a cube with fixed size in the anisotropic comoving frame, equivalent to a rectangular box with evolving (and unequal) side-lengths in the physical frame. 

Our simulations have three additional parameters in comparison to a standard cosmological N-body simulation. These are the three eigenvalues of the tidal field $\lambda_i$ which define the time evolution of the three axes $\alpha_i$ of the anisotropic universe as described in \ref{sec:background}. We have implemented the necessary changes into \gadget \hlt{\citep{gadget4paper}}, an updated version of \Gadget2 \citep{2005MNRAS.364.1105S}. 

The major required modifications are (1) the numerical integration of the anisotropic background scale factors; (2) changes to the equations of motion; (3) changes to the long- and short-range force-calculations. The equations relevant for (1) and (2) are summarized in section \ref{sec:background}--\ref{sec:equationsofmotion}; their implementation is described in our previous paper \citep{Schmidt:2018}. Therefore we will focus here on point (3) -- a description of the necessary changes to the force calculation in a TreePM scheme.

\subsection{The Tree-PM force split}
In \gadget the gravitational potential is split into a short-range part $\phi_s$ which is calculated by tree-summation techniques and a long-range part $\phi_l$ which is calculated on a periodic particle mesh through Fourier-techniques \citep{Bagla:2002,2005MNRAS.364.1105S}. Our calculation of forces in the anisotropic frame requires modifications to both of these.

The force-split is defined by a kernel function $f$ where the long-range potential is defined as a smoothed version of the potential and the short-range potential is defined as the remaining part of the potential so that $\phi = \phi_s + \phi_l$:
\begin{align}
    \phi_l &= \phi * f \,,\\
    \phi_s &= \phi  - \phi_l \,,
\end{align}
where the star denotes a convolution. 

In \gadget a Gaussian kernel is used for the smoothing function $f$,
\begin{align}
    f   &= \frac{1}{8 \pi^3 r_s^3} \exp \left(- \frac{|\myvec{x}|^2}{4 r_s^2} \right) \,,
\end{align}
whose Fourier representation (denoted by a tilde) is
\begin{align}
    \Tilde{f} &= \exp (- k^2 r_s^2)\,,
\end{align}
where $r_s$ is a parameter which defines the scale of the split and is typically chosen to be a bit larger than a mesh cell. We adopt this choice here for the anisotropic comoving frame; that is, $k$ and $\myvec{x}$ are given in anisotropic comoving coordinates so that the force-cut is spherical in the simulation frame. This gives a simple representation of the long-range force and a force-cut that does not deform in the simulation frame over time.  However, the kernel shape is ellipsoidal in the isotropic comoving frame which comes at the cost of a complicated form for the real-space representation of the short-range potential. The Green's function of the short-range potential requires the calculation of the potential of an ellipsoid with a Gaussian kernel. We will describe how to handle this in section \ref{sec:shortrangeforce}.

\subsection{Evaluation of the long-range force}
The Poisson equation for the long-range potential reads
\begin{align}
\sum \alpha_i^{-2} \partial_i^2 \phi_l &=  \frac{4 \pi G \rho_0 \delta}{\alpha_1 \alpha_2 \alpha_3} * f \label{eqn:poisson_long}
\end{align}
and can be solved in Fourier space as
\begin{align}
  \Tilde{\phi}_{l} &= - \frac{4 \pi G \rho_0 \Tilde{\delta}}{\alpha_1 \alpha_2 \alpha_3} \frac{\Tilde{f}}{ \sum \alpha_i^{-2} k_i^2 } \,.
\end{align}
Thus evaluating the long-range potential only requires a modified background density $\rho_0 / (\alpha_1 \alpha_2 \alpha_3)$ and a modified Greens's function in Fourier space,
\begin{align}
  \Tilde{G}_l = \frac{\tilde{f}}{\sum \alpha_i^{-2} k_i^2 } \,.
\end{align}
The rest of the calculation can be kept the same as in the isotropic case.
We have already presented this in \cite{Schmidt:2018} for the case $\Tilde{f} = 1$, since we used the code in a PM-only setting.

\subsection{Evaluation of the short-range force} \label{sec:shortrangeforce}

\label{sec:AnisoTree}
To estimate the Green's function of the short-range potential we have to find a real-space representation of the long-range potential:
\begin{align}
  G_s &= G - G_l \,, \\
  G_l &= G * f \,. \label{eqn:shortrangeconvolution}
\end{align}
We find it easier to solve this convolution in the isotropic comoving frame $r_i = \alpha_i x_i$ where the Green's function and the kernel are given by
\begin{align}
    G 
      &= \frac{1}{|\myvec{r}|} \,, \\
    f(u) &=  N \exp(- u^2 / (4r_s^2)) \,, \label{eqn:ellipsodialmass}\\
    u^2(\myvec{r}) &= \sum_i \frac{r_i^2}{\alpha_i^2} \,,
\end{align}
with $N=1/(8 \pi^3 r_s^3)$. We can then rephrase the convolution in \eqref{eqn:shortrangeconvolution} as the convolution of an ellipsoidal mass distribution with Gaussian kernel \eqref{eqn:ellipsodialmass} with the gravitational potential kernel $1/|\myvec{r}|$. The potential of an ellipsoid has already been solved in the literature \citep[e.g.][]{1969efe..book.....C, 2008gady.book.....B}. We show in Appendix \ref{sub:EllipticalP} that this is given for the Gaussian case by

\begin{align}
G_{l}(\myvec{r}) &= 2\pi \alpha_1 \alpha_2 \alpha_3 N\sigma^2 \nonumber \\
&\times \int_{0}^{\infty}\frac{\exp\left[-\frac{1}{2\sigma^2} \left(\left(\frac{r_1^2}{\alpha_1^2+v}\right) + \left(\frac{r_2^2}{\alpha_2^2+v}\right) + \left(\frac{r_3^2}{\alpha_3^2+v}\right)\right)\right]}{\left((\alpha_1^2+v)(\alpha_2^2+v)(\alpha_3^2+v)\right)^{1/2}} dv\,,
\label{eq:LongPot2}
\end{align}
with $\sigma = \sqrt{2} r_s$. We have not been able to find a closed-form expression for this integral.

However, since we only expect moderate axis ratios $0.5 \ll \alpha_i/\alpha_j \ll 2$, we can expand the solution as a perturbation to the spherical symmetric case $\alpha_i = \overline{\alpha}$ which can be solved analytically. We derive this in detail in Appendix \ref{sub:EllipticalP} and \ref{sub:EPA}. This expansion can be parameterized in the form
\begin{align}
G_{l}(\myvec{x}) &\approx 4\pi \alpha_1 \alpha_2 \alpha_3 \rho_{0}r_{s}^2 \nonumber \\
&\times\left(I_{3}(r) + \sum_{i} \left(I_{5}(r) \bar{\alpha} + I_{7}(r) r_{i}^2 \frac{\bar{\alpha}}{2r_{s}^2}\right)\overbrace{(\alpha_{i} - \bar{\alpha})}^{\equiv\Delta \alpha_{i}} + ...\right)\,.
\label{eq:phi_long}
\end{align}
where the different $I_{m}$ are defined in \eqref{eq:Im} and are only a function of the radius $|\myvec{r}|$. We give the full expansion of the potential up to second order in $\Delta \alpha_i$ in Appendix \ref{sec:AppendixPot}.

We show in Appendix \ref{app:approximation_error} that the error of this expansion leads leads to extremely small force errors for axes that all deviate slightly from unity, and the errors become large when the relative differences between the axes reach order unity. For typical cases that we consider in this paper we have $\lambda \sim \Delta \alpha \sim 0.1$ and the largest relative force errors are of order $10^{-3}$. We estimate that the relative force errors still stay below $1\%$ for cases up to $\Delta \alpha \approx 0.4$. The errors can be seen as a function of the axes in Figure \ref{fig:forceerrorvsalpha}.

The short range force is then estimated in the simulations by doing a tree-summation over the short-range force of a point-mass given by $\myvec{\nabla} G_s$ instead of the spherically symmetric short-range force. The rest of the Tree-PM algorithm can be kept the same as in \cite{2005MNRAS.364.1105S}. However, we note that the tree is grouping particles in cubes in the anisotropic frame, leading to cuboids in physical space. In general there is nothing which speaks against such a different grouping. However, in our simulations we use smaller opening angles and more accurate force-accuracy parameters to make sure that the force accuracy does not suffer from this different grouping strategy.

In addition to the tests in Appendix \ref{app:approximation_error}, we have also tested our anisotropic force calculation for an evolved particle distribution against a cuboid mesh implementation in the unmodified \gadget -- resembling an anisotropic box with non-evolving axes. We found force errors which are consistent with our measurements in Appendix \ref{app:approximation_error}.

We conclude that the force calculation presented here is suitable for performing high-accuracy cosmological simulations with a large-scale tidal field.

\section{Response Measurement and numerical Convergence}
We have discussed in the last sections how to perform simulations in an anisotropic universe. We will describe here how to measure the effects of the tidal field. These can be quantified at the first relevant order by the response, which describes the development of an anisotropic component in the power spectrum. Here we briefly summarize the definition of the response, we explain how to measure it in a simulation, and we discuss its convergence within our simulations.
\label{sec:Response}
\subsection{Response Definition}
Our response definition follows that of \cite{2017JCAP...06..053B}. In this paper we will focus on the first-order response functions as set out in their Sec. 3.2. Under the influence of an external \emph{trace-free} tidal field $K$ and a large scale overdensity $\delta_{L}$, the three-dimensional power spectrum\footnote{By this we mean $P(\myvec{k}) = \langle |\Tilde{\delta}^2(\myvec{k})|\rangle$ where the average is over realisations of the linear initial conditions.} can be written in the anisotropic comoving frame as
\begin{equation}
P(\myvec{k}) = P(k) \left(1+ G_{1}(k)\delta_{L} + G_{K}(k) \sum_{i,j} \hat{k}_{i}\hat{k}_{j}K_{ij}\right)\,, 
\label{eq:gkresponse}
\end{equation}
where $\hat{\myvec{k}}$ is a normalized $\myvec{k}$-vector $\hat{\myvec{k}} = \myvec{k}/k$ and $K$ the traceless tidal tensor of the large-scale perturbation. In the linear regime $K$ is given in its eigenframe by
\begin{align}
    K(t) = D(t)
\begin{pmatrix}
\lambda_1 - \delta_L/3 & 0 & 0 \\
0 & \lambda_2 - \delta_L/3 & 0 \\
0 & 0 &\lambda_3 - \delta_L/3
\end{pmatrix}\,,\label{eqn:Kaslambda}
\end{align}
where $\delta_L = \sum \lambda_i$. Equation \eqref{eq:gkresponse} remains valid for small $\delta_L$ and $K_{ij}$ and as long as the wavelength of the large-scale density and tidal perturbations is much larger than $1/k$. In our case the wavelength is infinite, so that this condition is satisfied for all $k$.

\cite{2017PhRvD..95h3522A,2017JCAP...06..053B} found to leading order in perturbation theory that
\begin{equation}
G_{K} = 8/7\,, \label{eqn:responseedsa}
\end{equation}
which is valid on large scales as $k \rightarrow 0$ for an Einstein-de-Sitter universe. We show in Appendix \ref{app:response} that in a $\Lambda$CDM universe a more accurate approximation is given by
\begin{align}
 G_{K} \approx \frac{8}{7} \Omega_m^{1/185} (a)\,,
\end{align}
which, however, deviates by less than a percent from equation \eqref{eqn:responseedsa} at $a=1$. As we will describe in the next section, it is straightforward to measure $G_{K}$ from our simulations in the anisotropic frame. However, in addition to $G_K$, it is of interest to measure the response in the isotropic frame, $R_K$. This follows an equivalent definition to \eqref{eq:gkresponse} but in the isotropic comoving Fourier space, and can be inferred by applying a coordinate transformation to $G_K$,

\begin{equation}
R_{K} = G_{K} - k \frac{P'(k)}{P(k)}\,, \label{eqn:responsetrafo}
\end{equation}
where $P'(k) = dP(k)/dk$ is the derivative of the background (isotropic) power spectrum.




\subsection{Response Measurement} \label{sec:responsemeasurement}

With our setup we can simulate universes with arbitrary large-scale over-densities and/or tidal fields. Therefore we can measure the response by performing simulations in universes with small differences in the large-scale tidal field -- parameterized by $\myvec{\lambda}$. Using $\eqref{eqn:Kaslambda}$ and assuming a trace-free perturbation $\sum \lambda_i = 0$, we can write for universes with small values of the tidal field
\begin{align}
  P(\myvec{k}, \myvec{\lambda}) &\approx P(\myvec{k}, \myvec{\lambda}=0) \left(1 +  G_K \sum \lambda_i \hat{k}_i^2 \right)\,.
\end{align}
This motivates us to infer the response by performing triplets of simulations which start from the same initial condition realization, but different tidal fields: two with opposite signs $\pm \myvec{\lambda}$ of the tidal field and one without a tidal field $\myvec{\lambda}=0$. Then the response can be measured by a finite-difference scheme of the form
\begin{align}
  G_K &=  \frac{P(\myvec{k}, \myvec{\lambda}) - P(\myvec{k}, -\myvec{\lambda})}{P(\myvec{k}, \myvec{\lambda}=0) \sum 2 \lambda_i \hat{k}_i^2 } + O(|\myvec{\lambda}|^2)\,. \label{eqn:symresponse}
\end{align}


At linear order in the tidal field, we can specialise to the case without an overdensity $\sum_i \lambda_i = 0$ and with a tidal field which is axisymmetric, $\lambda_x = \lambda_y  = - \lambda_z/2$. This way we can reduce the $\lambda_i$ to a single parameter $\lambda_z$.
In this case we have
\begin{align}
\sum \hat{k}_i^2 \lambda_i &= \frac{\lambda_{z}}{2} \left(3\hat{k}_{z}^2-1\right) \nonumber \\
&= \lambda_{z} Y_{2}(\mu) \,.
\label{eq:tidal_special_case}
\end{align}
with $Y_2$ the second-order Legendre polynomial, and $\mu = \hat{k}_{z}$, the cosine of the angle between the k vector and the z axis in Fourier space. 
In principle, equation \eqref{eqn:symresponse} should be valid for each mode $\myvec{k}$, since it is already defined over expectation values. However, with a single initial condition realisation simulated three times, we cannot determine these expectation values reliably for individual modes, but rather must perform some sort of averaging over an ensemble of Fourier modes. We found it elegant to spherically average the following expression:
\begin{align}
    G_K &= \frac{\langle (|\delta^2_+| - |\delta^2_-|) Y_2(\mu) \rangle }{\langle  |\delta^2_0|  Y_2^2(\mu) D(t) \lambda_z \rangle} + O(\lambda_z^2)\,, \label{eqn:responseaverage}
\end{align}
where $\delta_{\pm} := \delta(\myvec{k},\pm\lambda_z)$. 
The weighting by $Y_2$ extracts the response information optimally, and we note that, due to the spherical symmetry of the run without tidal field,  the denominator should factorize approximately into
\begin{align}
   \langle  |\delta^2_0|  Y_2^2(\mu) D(t) \lambda_z \rangle &= \langle  |\delta^2_0| \rangle \langle  Y_2^2(\mu) \rangle D(t) \lambda_z \\
    &= P_0 D(t) \lambda_z\,,
\end{align}
where we used the normalization condition for a Legendre polynomial.

This strategy allows us to measure the response as a function of wavenumber $k$ on a grid in Fourier space. We employ folding techniques to measure the power spectrum and the response on very small scales. We describe the numerical details of this in Appendix \ref{app:response_measurement} where we also investigate the effects of other numerical parameters.

We note that the error in the response measurement \eqref{eqn:responseaverage} scales as $\lambda_z^2$. We typically employ simulations with $\lambda_z \sim 0.1$ leading to errors which are of order a percent in the response measurement.

\subsection{Time evolution and discreteness Effects}
\begin{figure*}
    \centering
    \includegraphics[width=0.9\textwidth]{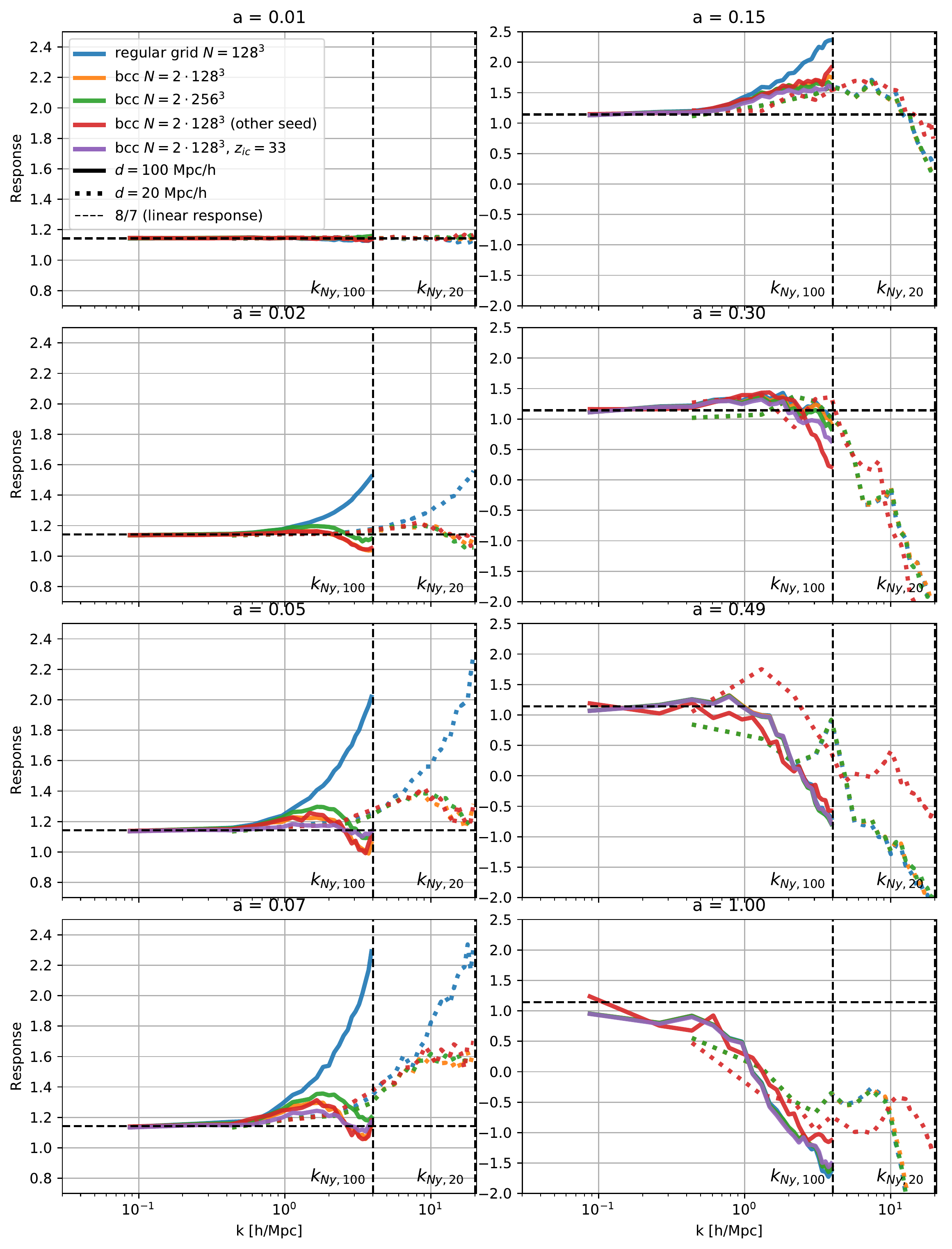}
    \caption{Time evolution and convergence tests of our response measurements. Different panels correspond to  different times as labeled. Dashed lines refer to simulations with boxsize $d = \SI{20}{\per\h\mega\parsec}$, while solid lines are for $d = \SI{100}{\per\h\mega\parsec}$. Blue lines correspond to simulations where the initial particle load was a cubic grid. The other runs use a bcc grid, as described in the text. Orange and green lines differ by a factor of eight in the number of particles used. The purple line refers to a later start, $z_{ic}=33$ rather than 99, as in all the other simulations. All but the red simulations start from the same initial condition realisation, and at later times agreement between them is often so good that only the last one plotted is visible, even though all simulations are, in fact, shown at all times. The \hlt{simulations shown as red and orange lines (note that the orange line is mostly covered by the green line)} are identical except for using different seeds in the random number generator; differences between them thus indicate the effect of cosmic variance. At $a=0.3$ and later, differences between runs can be explained solely by cosmic variance.}
    \label{fig:early_response}
\end{figure*}

We begin by analysing a set of test runs to check the reliability of our response measurement and  simulation techniques. These were carried out using box sizes of $d = \SI{100}{\per\h\mega\parsec}$ and $\SI{20}{\per\h\mega\parsec}$, and with different resolutions, initial grid discretisations and starting redshifts ($z_{ic} = 99$ except for one run with $z_{ic} = 33$). Results are presented in Figure \ref{fig:early_response} with different panels showing the response measured at different times. 

In our initial conditions, our response estimate is exactly that predicted, but at subsequent early times ($0.01<a<0.3$) it is artificially enhanced in a way which depends on the particular discretisation pattern chosen for the initial particle load. At later times  ($a>0.3$) this artifact disappears and there is good agreement for all tested choices of numerical parameters (up to differences which can be explained by cosmic variance). A more detailed explanation follows.
  
The blue lines in Figure \ref{fig:early_response} correspond to our default setup, a total of $N= 128^3$ particles initially displaced from a cubic grid. The initial conditions ($a = 0.01$) show exactly the response expected according to linear theory, confirming the validity of our initial conditions generator. However, as evolution starts, the response increases at wavenumbers above a threshold which scales inversely with the mean interparticle  separation (compare the $\SI{20}{\per\h\mega\parsec}$ and $\SI{100}{\per\h\mega\parsec}$ runs). This growth seems to be a numerical artifact, and depends strongly on the discretisation pattern used to arrange particles in the initial conditions prior to imposing Lagrangian perturbation theory in what is also sometimes called ``pre-initial conditions''. 


The effect of the discrete particle distribution on the growth of modes at the smallest scales can be understood by comparing our setup to one where initially a body-centered cubic (bcc) lattice is used as the \hlt{unperturbed particle distribution}. This is known to have less anisotropic growing modes induced by discreteness effects \citep[cf.][]{Joyce2007,Marcos2008}. \hlcom{Added this sentence:} For details on the set-up and effect of pre-initial conditions, we kindly refer the reader to \cite{Michaux:2020}, in particular their Section~2.4.  We find that the artificial early growth is very significantly suppressed in this case. There is still some apparently unphysical early enhancement of the response, but this does not disappear or move to significantly smaller scales if the mesh spacing of the initial load is halved (the green lines) or if a later starting redshift ($z=33$) is used (the purple lines). We have run numerous other numerical tests, but we were unable to find a simple consistent explanation of this early growth. At least the large effect seen in the blue curves is clearly a consequence of the details of the particle pattern in the initial load, and so must be a discreteness artifact.

A reliable estimate of the response at high redshift $z \gg 2$, will require a full identification and proper resolution of the sources of these problems -- this may be possible using a phase-sheet-based approach as described in \citet{hahn_angulo_2016} and \citet{stuecker_2019} -- but Figure \ref{fig:early_response} demonstrates that although our runs show a variety of anomalies at early times, $a \ll 0.3$, they agree with each other remarkably well for $a \gtrsim 0.3$. Indeed, by $a=1$ differences are very small and many of the curves of Figure \ref{fig:early_response} overlie each other. (Note that they were plotted in the order they are listed in the legend, with later curves overlying earlier ones.) Apparently, the early artificial growth of the response (which differs between runs) has little impact on its value at later times. This is probably because the late-time behaviour is dominated by physical processes like halo formation which are insensitive to details of the linear evolution.  Remaining differences at $a \gtrsim 0.3$ can be attributed to cosmic variance between the simulations. To demonstrate this, we ran simulations with a different seed in the random number generator for the initial conditions generator; these are plotted as the red lines (to be compared to the orange lines) in Figure \ref{fig:early_response}. At late times the scatter between red and orange is \hlt{roughly comparable} to the difference between the smaller and larger boxes ($d = \SI{100}{\per\h\mega\parsec}$ and $d = \SI{20}{\per\h\mega\parsec}$). \hlcom{Added this sentence:} There might of course still be systematic differences due to finite-size effects in the simple tests in this section, but we will use simulations with much larger boxsize and multiple different seeds in the following analysis.

We conclude that our setup is able to measure the response reliably at late times $a \gtrsim 0.3$, where it seems to be robust against  discretisation details. However, a precise measurement at earlier times would require more sophisticated techniques. We therefore focus in the rest of this paper on a precise measurement of the response for $0\leq z\leq 2$.

\subsection{Production Simulation Setup}
\label{sec:Simulations}
To measure the response on both large and small scales, we perform a set of cosmological gravity-only simulations with side length $L_{B} = \SI{500}{\per\h\mega\parsec}$ with  $512^3$ particles each. This corresponds to a particle mass of $\SI{8.0e10}{\per\h} M_{\odot}$. These simulations use initial conditions as described in section \ref{sec:initialconditions}, and are evolved with the modified \Gadget 4 version described in section \ref{sec:modGadget}. The gravitational force is calculated using the modified TreePM algorithm using the elliptical potential approximation and $1024^3$ PM cells.

We consider 8 realisations of the initial density and velocity field at our starting redshift $z_{\text{ini}} = 127$, but for each we perform three runs with different tidal fields defined by $\lambda_z = -0.1$, $\lambda_z = 0$ and $\lambda_z = 0.1$ as explained in section \ref{sec:responsemeasurement}. \hlcom{Changed this sentence:} We use these 8 realizations to measure the average amplitude of the response and the statistical uncertainty of the average. 

The inital power spectrum is computed with \textsc{camb} \citep{Lewis:1999bs} using a fiducial flat $\Lambda$CDM cosmology with cosmological parameters taken from Planck \citep{Planck2015}.
The main parameters are $\Omega_{m} = 0.308$, $\Omega_{\Lambda} = 0.692$, $\Omega_{b} = 0.04694$, $\sigma_{8} = 0.829$, $n_s = 0.965$ and $h = 0.678$. The fiducial runs use a softening length of $\SI{40}{\per\h\kilo\parsec}$ which translates to $0.04 \cdot l_{\mathrm{mean}}$ with the mean particle separation, $l_{\mathrm{mean}} = \SI{500}{\per\h\mega\parsec} / 512 \approx \SI{0.98}{\per\h\mega\parsec}$.
For convergence tests we also ran simulations both with half that softening and with one eighth the particle number and the same softening. 

We have carried out additional tests of the convergence of our simulations on small scales, and of the way numerical details affect our response measurements. We present these tests in Appendix \ref{app:response_measurement} with a detailed explanation, and we show in Figure \ref{fig:response_measurement} how power spectra, and the responses $G_{K}$ and $R_{K}$ depend on numerical details. \hlcom{Weakened the statement of this sentence:} As discussed in Appendix  \ref{app:response_measurement}, we estimate that the response measurements in our simulations are reliable up to $k = \SI{7}{\h\per\mega\parsec}$. 

\section{Results}
\label{sec:Results}
In this section we describe the main results of this paper. These are a measurement of the power spectrum response on small scales, up to \hlt{$k = \SI{7}{\per\h\mega\parsec}$}, and a first look at the effect of the tidal field on the alignment of haloes.

\subsection{Response Function}
\label{sub:Response_result}


\begin{figure*}
    \centering
    \includegraphics[width=0.8\textwidth]{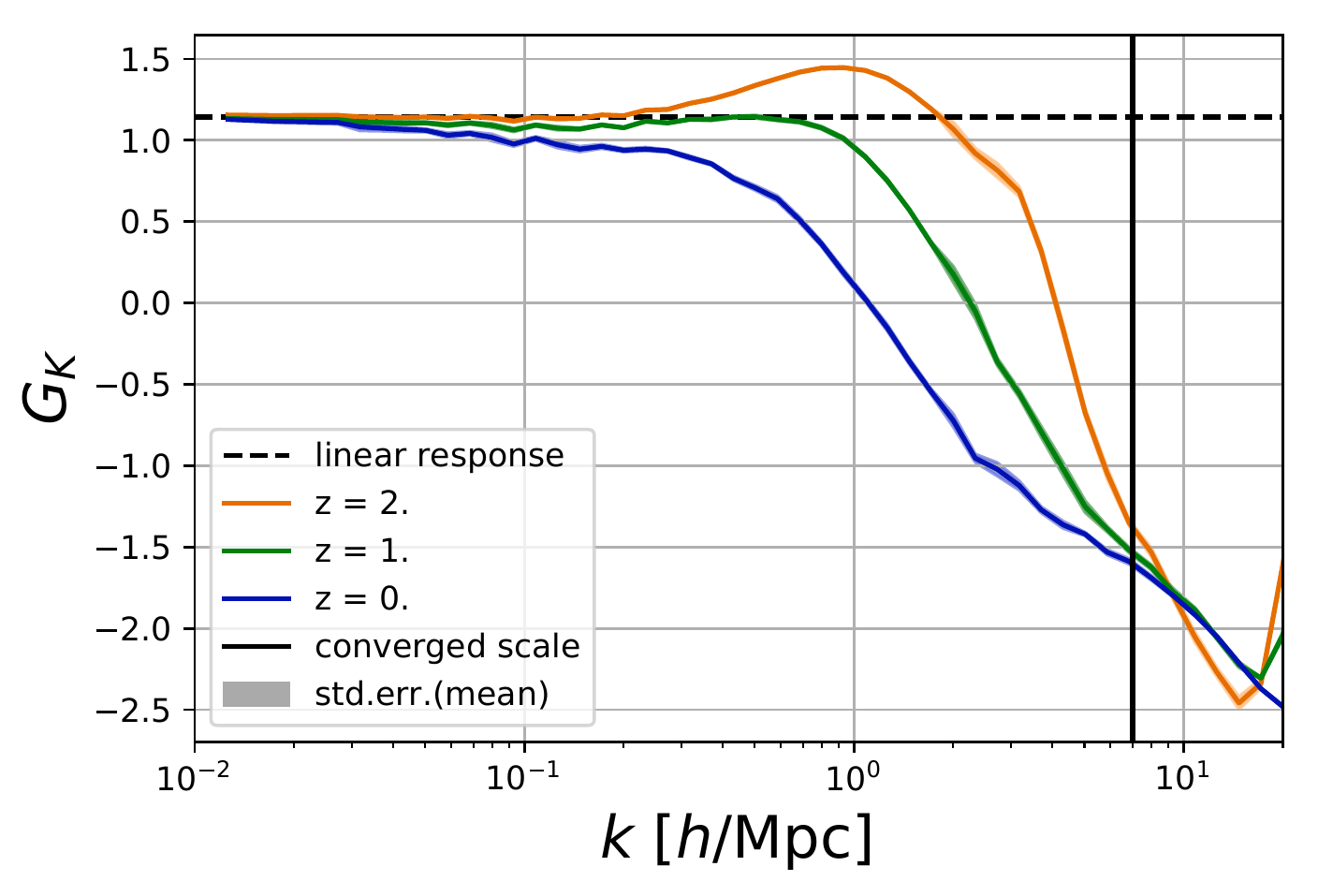}
    \includegraphics[width=0.8\textwidth]{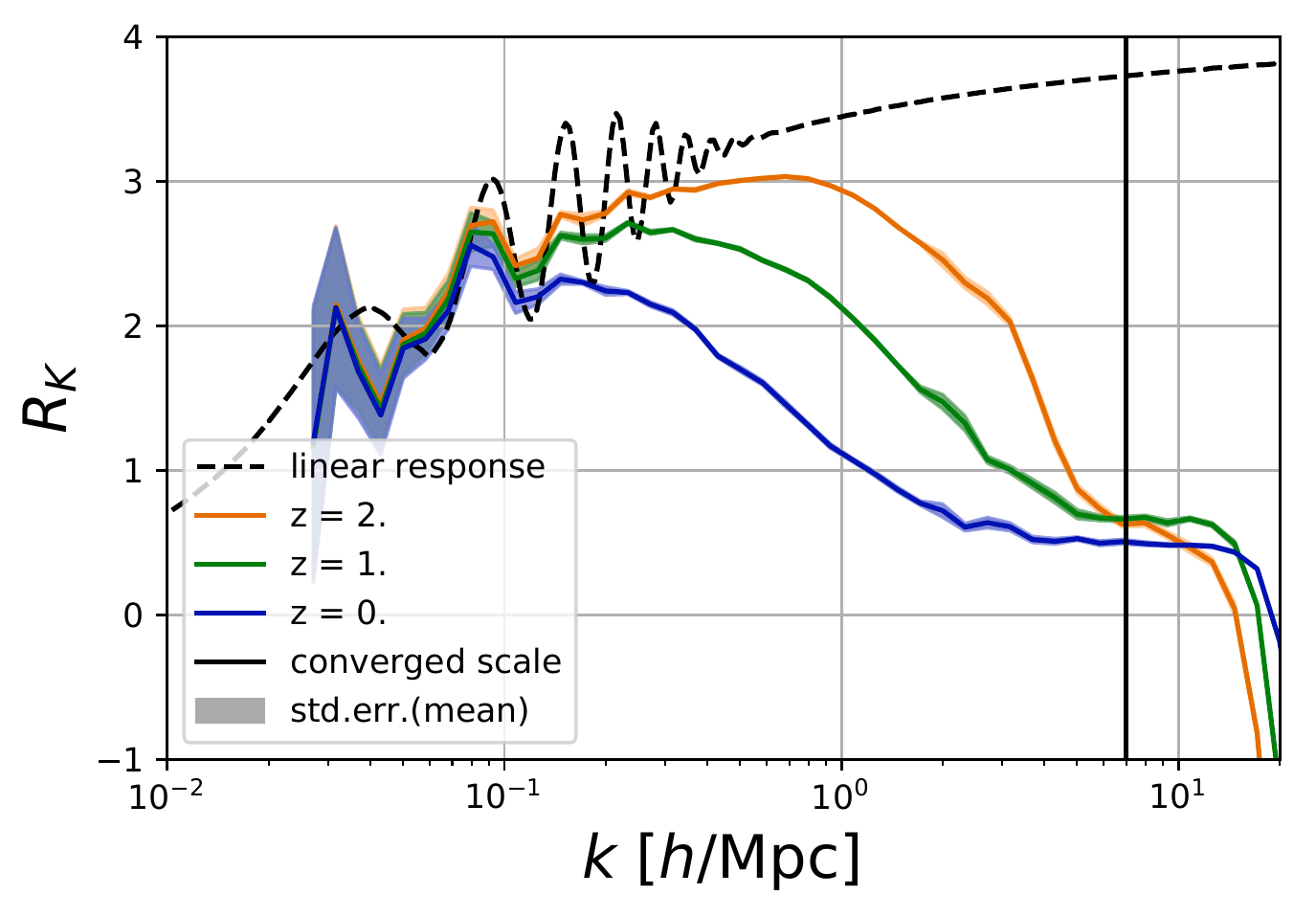}
    \caption[Response function - TreePM]{\hlcom{Moved vertical line to $k = 7$Mpc$/h$}Response functions for three different redshifts. The response in the anisotropic comoving frame (``growth-only response'') $G_{K}$ (top) approaches the linear limit $G_{K} \approx 8/7$ on large scales and shows mostly a suppression on smaller scales. The length-scale of the suppression grows with time. However, at $z=2$ one can also see a slight enhancement over the linear prediction at scales around $k \sim \SI{1}{\h\per\mega\parsec}$ which are just above the suppression scale. The response in the isotropic comoving frame (``total response'') $R_{K}$ (bottom) shows an oscillatory behaviour at large scale which is caused by the shift in the baryon acoustic oscillation feature. Note that this feature is smoothed out through the binning procedure, in addition to the smoothing due to nonlinear evolution. At smaller scales $k \gg \SI{0.1}{\h\per\mega\parsec}$ the response shows a redshift-dependent suppression with respect to the linear prediction. At very small scales $k \gtrsim \SI{2}{\h\per\mega\parsec}$ it approaches a value of $\simeq 0.5$.}
    \label{fig:new_response_treepm}
\end{figure*}

Based on the results shown in Figure \ref{fig:early_response}, the deviations in response measurements after $a=0.3$ appear to be due primarily to cosmic variance, with numerical uncertainties being relatively small. In Figure \ref{fig:new_response_treepm} we therefore present the  mean response functions $G_{K}$ and $R_{K}$, together with the error on the mean obtained from the scatter among our 8 realisations, at redshifts $z=0$, $1$ and $2$ for which we believe discreteness artifacts to be small. 

We find the linear predictions for the response functions to be well reproduced on large scales, but the response is suppressed on quasi-linear and nonlinear scales. As the characteristic (comoving) scale of nonlinearity  grows with time, so does the scale on which the response is suppressed below the linear value; i.e., the suppression sets in at lower wavenumbers at later times. 

Interestingly, the response $R_{K}$ in the isotropic comoving frame (``total response'') approaches a constant value $R_{K} \approx 0.5$ on small, highly nonlinear scales, which appears to be redshift-independent. As discussed in detail in Sec.~5 of \cite{Schmidt:2018}, in the context of a halo model description of the nonlinear power spectrum, this implies that the tidal field has an impact even on the innermost structure of haloes and down to low halo masses. If the inner structure of haloes were insensitive to the tidal field, one would expect $R_K = 0$ at large wavenumbers. We investigate the effect on haloes more directly in the next section, although the precise connection between halo alignments and the tidal matter power spectrum response on small scales is left for future study.



\subsection{Effect of the tidal field on halo alignments}

\begin{figure}
    \centering
    \includegraphics[width=0.48\textwidth]{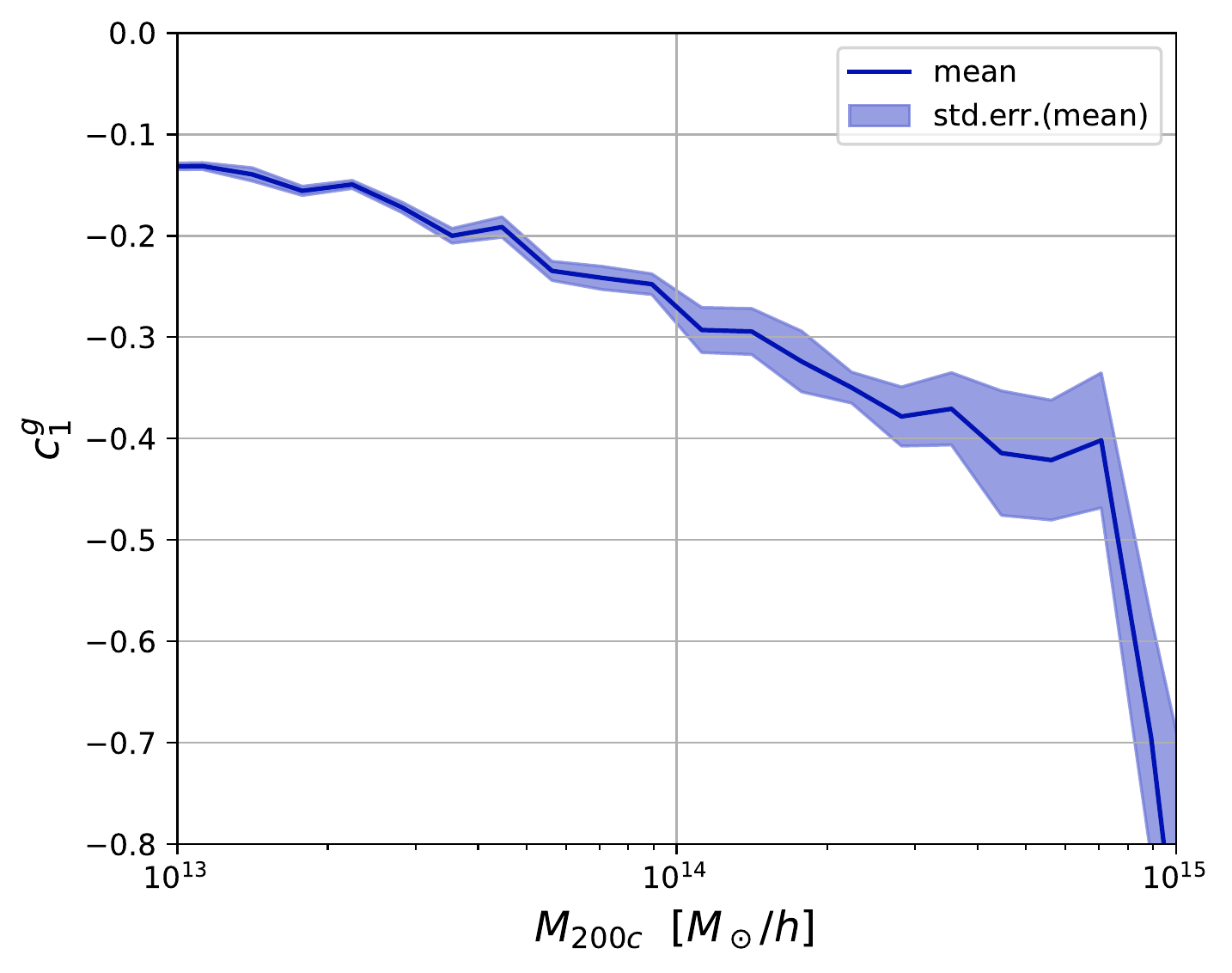}
    \caption{Alignment coefficient for the reduced moments of inertia tensor of haloes at $z=0$, as measured using equation \eqref{eqn:c1g} in logarithmic bins in halo mass with width $\Delta \log_{10} M_{200c} = 0.1$. The shaded band indicates the error on the mean as determined from the 8 simulation realizations. The alignment with the tidal field is stronger for haloes with larger masses. Note that the alignment coefficient $c_{1,g}$ is negative, since the inertia tensor is increased in the direction of negative $\lambda$ (a stretching tidal field).}
    \label{fig:c1g}
\end{figure}

As just discussed, our response measurements suggest that tidal fields influence even the innermost structure of haloes. In this section we investigate this effect more quantitatively. A proper treatment of such tidal alignment is important for analyses of weak gravitational lensing, where it can lead to significant contamination of shear estimates.

We adapt the {\small  SUBFIND} algorithm \citep{springel2001} to identify haloes in our simulation in isotropic comoving coordinates (as usually done, although in these coordinates our simulation boxes are not cubic). We compute the reduced inertial tensor,
\begin{align}
  I_{ij} = \frac{1}{N} \sum_n \frac{r_{n,i} r_{n,j}}{r_n^2}\,,
\end{align}
where $\myvec{r}_n$ denotes the displacement vector between the particle with index $n$ and the potential minimum of the halo. Thus $I_{ij}$ measures shapes as viewed in the isotropic background universe. The sum extends over all particles within the radius $R_{\text{200c}}$ of a given halo, where $R_{\text{200c}}$ is defined such that the mean matter density inside $R_{\text{200c}}$ is 200 times the critical density of the universe. Therefore $I_{ij}$ is an indicator for the shape and orientation of a halo in the physical frame.  We have tested different radius and mass definitions, in addition to the regular, un-reduced inertia tensor, and found that all cases give similar results on the alignments. We note that more sophisticated algorithms to estimate the shapes of haloes exist (see \cite{alignment_review} for a review), but this one suffices for our purposes.

  Perhaps the simplest way to measure halo alignments is to average the inertial tensor over many halos in a fixed coordinate frame (in our case the isotropic comoving frame). In the absence of any preferred directions, the average inertial tensor is proportional to the identity matrix $\delta_{ij}$. The external tidal field $K_{ij}$ however provides a preferred direction along which halos can align, so that to linear order in $K_{ij}$ (or equivalently $\lambda$) we can write
\begin{align}
  \langle I_{ij}\rangle_\text{haloes} \Big|_\lambda = \frac13 I \left[ \delta_{ij}
    + c_1^{\rm g} K_{ij}\right],
  \label{eqn:Iexp}
  \end{align}
where $I = {\rm tr}[I_{ij}]$ and we have introduced a dimensionless fractional alignment coefficient $c_1^{\rm g}$ following the notation of \cite{vlah/etal}.
Equation \eqref{eqn:Iexp}, is in fact the leading contribution in the general perturbative expansion of intrinsic alignments \citep{Catelan01,blazek/etal,schmidt/chisari/dvorkin,vlah/etal}. It is the analogue for shapes of the linear galaxy $b_1$, and so describes the statistics of three-dimensional shapes in the large-scale limit. The coefficient $C_1$ of the relation between projected shapes and the tidal field in the linear-alignment model (e.g., \cite{blazek/etal:2011}) is related to $c_1^{\rm g}$ via $C_1 = -c_1^{\rm g} D(a)/(a^2 \rho_m(a))$, so that $c_1^{\rm g} < 0$ corresponds to $C_1 > 0$.

The trace $I$ of the inertial tensor is not affected by the tidal tensor at linear order, and can be straightforwardly measured by averaging over halos in the $\lambda_z=0$ simulations. Since in our simulations $K_{ij} = {\rm diag}(\lambda_z/2, \lambda_z/2,-\lambda_z)$, one can directly solve for the alignment coefficient in terms of components of the average inertial tensor in the simulations with tidal fields:
\begin{align}
c_1^{\rm g} = \frac{{\rm TF}[I_{ij}]_{+\lambda_z} - {\rm TF}[I_{ij}]_{-\lambda_z}}{I \lambda_z}\,, \label{eqn:c1g}
\end{align}
where
\begin{align}
  {\rm TF}[I_{ij}]_{\lambda_z} =  \left( \frac12 \langle I_{xx}\rangle_\text{haloes}
  + \frac12 \langle I_{yy}\rangle_\text{haloes}
  -\langle I_{zz}\rangle_\text{haloes} \right)\Big|_{\lambda_z}\,.
\end{align}
The results of the measurement are shown in Fig.~\ref{fig:c1g}.
As expected, $c_{1}^{\rm g}$ is negative, since the inertia tensor is increased in the direction of negative $\lambda$ (a stretching tidal field).
We find that more massive halos are aligned more strongly, in agreement with previous findings \citep{Jing:2002mm}.

\begin{figure}
    \centering
    \includegraphics[width=0.48\textwidth]{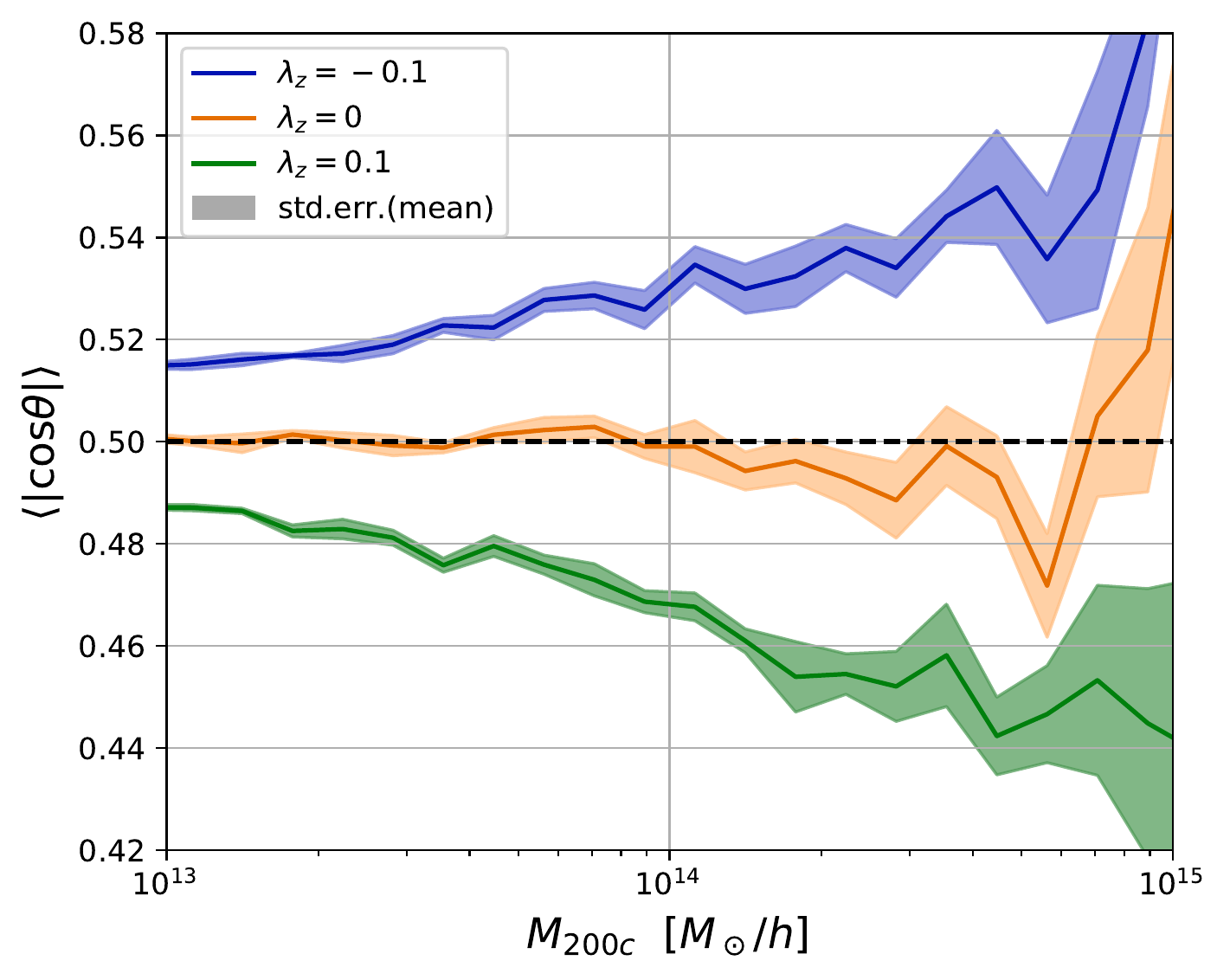}
    \caption{Mean alignment $\langle |\cos \Theta| \rangle$ of haloes at $z=0$, in the same mass bins as Fig.~\ref{fig:c1g}, with the major axis of the tidal field in different mass bins. The shaded contours highlight the uncertainty in the mean as determined from the different simulation realizations. The results for the isotropic run without tidal field are consistent with a mean of 0.5 at all masses, which corresponds to the expected uniform distribution of angles. For nonzero $\lambda$, haloes tend to align with the ''stretching'' direction of the tidal field. We again find that haloes with higher masses align more strongly.}
    \label{fig:mean_angle}
\end{figure}

As another measure of alignments, we determine the eigenvalues and eigenvectors of the reduced inertia tensor and compute the angle $\theta$ between the direction in which the tidal field is strongest (the $z$-axis) and the eigenvector associated with the largest eigenvalue of the reduced inertia tensor. We plot the mean value of $|\cos(\theta)|$ in Figure \ref{fig:mean_angle} for our three cases with differing tidal fields. For random alignments, the expectation  is $\langle |\cos(\theta)| \rangle = 1/2$. Note that this measure does not depend on the axis ratios of the haloes, but only on the orientation of their major axes.
Quantitatively, we find that the mean of $\cos\theta$ depends linearly on the tidal field,
\be
\langle |\cos(\theta)| \rangle_\lambda - \frac12 \simeq - (0.2 - 0.4) \lambda\,,
\ee
again with stronger alignment seen at higher halo mass. 
The case with negative $\lambda_z$ corresponds to a tidal field which ``stretches" the matter distribution along the z-axis, while positive $\lambda_z$ corresponds to a compressive field along this same axis. Figure \ref{fig:mean_angle} thus shows that the longest axis of a halo tends to align with the direction of tidal field stretch, which is essentially the same effect as quantified by the parameter $c_1^{\rm g}$.  This effect can also be seen in Figure \ref{fig:halo_vis} which shows images of the same halo in the three different cases. 

Our results for this simple measure of halo alignment are in qualitative agreement with previous findings \citep{hopkins_2005, chen_2016}, whereas a quantitative comparison is difficult, because of the different techniques to define the statistics. However, we note that our approach makes it possible to draw clear causal relations. For example a simple explanation for the alignment of haloes suggests that haloes align because they are part of the same cosmic web-structures like filaments or pancakes \citep{hahn_2007,alignment_review}. However, in our simulations we clearly see that haloes can even be aligned if they are part of completely separate structures -- as long as they share a common large scale tidal field. 

Further our measurements have a high signal-to-noise, given our relatively small simulation volume. This suggests that anisotropic N-body simulations of the kind presented here are well suited for precision studies of halo alignment using more sophisticated halo shape and alignment estimators than employed in this first exploration.

\section{Conclusions}

In this paper we have described how to perform simulations in an anisotropically expanding universe. This technique simulates structure formation in the presence of a tidal field of very long wavelength that is effectively uniform across the simulation volume. We have shown how to set up initial conditions for such simulations by correctly taking into account second-order Lagrangian perturbation theory in the anisotropic comoving frame. Further, we have shown how to adapt the TreePM algorithm of standard cosmological simulation codes to carry out calculations in this anisotropic frame. We have then carried out the first high-resolution simulations that consistently include the large-scale tidal field from very early until late times, and from large to small scales.

We have found that discretisation effects play a role in such simulations on small scales and at early times, but they vanish at late times, $a \gtrsim 0.3$ for our simulations. As a result, we were able to predict power spectrum response functions for redshifts $z\leq 2$. The response we find agrees with perturbation theory on large scales at all redshifts, but is generally suppressed on smaller scales where nonlinear effects become important. Around $z \sim 2$, however, we can see a slight rise above the linear prediction at intermediate scales, which we believe to be physical. It would be interesting to compare this with a higher-order perturbative calculation.

The suppression of the response on small scales can be interpreted as arising because haloes are ``more spherical" than linear predictions would suggest. However, both indirect and direct evidence show that they are still affected by the tidal field. We find that the response $R_K$ converges to a nonzero value ($R_K \simeq 0.5$) on small scales. In the context of the halo model, this implies that halo shapes respond to the tidal field.

In addition, by measuring both the linear alignment coefficient $c_1^{\rm g}$ as well as the mean alignment between the major axis of haloes and the preferred axis of the external tidal field, we find that haloes are preferentially oriented along the direction of maximum tidal stretch. These results, as well as the trend with halo mass, confirm previous measurements in the literature, but differ in that they isolate the effect of large-scale tidal fields, and provide a very high signal-to-noise measurement.

We have used very simple estimators of halo shape and alignment here, so the very high signal-to-noise of our detection indicates the considerable potential of the anisotropic N-body simulation technique for studying such effects.
Further interesting applications include the response of the halo (as opposed to matter) power spectrum, and measurements of halo tidal bias.

\section*{Acknowledgements}
The authors thank Raul Angulo for helpful discussions and Volker Springel for help with  \gadget and helpful discussion about the convergence. We further thank Kazuyuki Akitsu for discussions and pointing out a typo in \cite{Schmidt:2018}, as well as Elisa Chisari for helpful hints on halo alignments.
JS acknowledges the support by Raul Angulo's ERC Starting-Grant 716151 (BACCO). AS is supported by DFG through SFB-Transregio TR33 ``The Dark Universe''.
FS acknowledges support from the Starting Grant (ERC-2015-STG 678652) ``GrInflaGal'' from the European Research Council. OH acknowledges funding from the European Research Council (ERC) under the European Union's Horizon 2020 research and innovation programmes, 
Grant agreement No. 679145 (COSMO-SIMS).

\section*{Data Availability}
\hlcom{Added this section}
The code for initial condition generation that has been extended and used in this study is publicly available at \url{https://bitbucket.org/ohahn/monofonic}. The modified \gadget simulation code and the data underlying this article will be shared on reasonable request to the corresponding author or to Fabian Schmidt\footnote{E-mail: fabians@mpa-garching.mpg.de}.



\bibliographystyle{mnras}
\bibliography{ref_paper} 



\appendix

\section{Derivation of the Anisotropic 2LPT solution} \label{lpt:appendix}

We show here the full derivation of the 2LPT solutions that are presented in section \ref{sec:initialconditions} and we present a more detailed discussion of the approximations involved.

We start from the equations of motion as already stated in equation \eqref{eqn:eqofmotionsingle}:
\begin{align}
  \myvec{\nabla} \left( \ddot{\myvec{x}} + 2 H^*(a) \dot{\myvec{x}} \right)
                              &= \hlm{-} \frac{3 H^2 (a) \Omega_m(a)}{2 \alpha_1 \alpha_2 \alpha_3} \delta(\myvec{x})\,. \label{eqn:equationofmotionapp}
\end{align}
 We then follow the standard 2LPT procedure as given in 
 \citet{Jeong_2010} and we refer the reader to this reference for a complete understanding of the individual steps. 

\subsection{Lagrangian Perturbation Theory}
By defining the Lagrangian displacement field $\myvec{\hlm{\Psi}}$ through
\begin{align}
    \myvec{x}(\myvec{q}, t) &= \myvec{q} + \hlm{\myvec{\Psi}} (\myvec{q}, t)\,,
\end{align}
and noting that the density contrast can be written\footnote{In the single-stream regime, i.e. before any shell crossing has happened.} as
\begin{align}
    \delta = \frac{1}{\det (\frac{\partial \myvec{x}}{\partial \myvec{q}})} - 1\,,
\end{align}
we can derive the master equation of Lagrangian perturbation theory
\begin{align}
    J \left(\delta_{ij} + \Psi_{i,j} \right)^{-1} \left( \psidd_{i,j} +  2 H^*_{ik} \psid_{k,j} \right) &= \frac{3}{2} \frac{H^2 \Omega_m }{\alpha_1 \alpha_2 \alpha_3} (J - 1)\,, \label{eqn:lptmaster}
\end{align}
where $\Psi_{i,j} = \partial_i \Psi_j$ is the partial derivative of the displacement field $\myvec{\Psi}$, and we imply summation over identical indices. $J$ is the determinant of the Jacobian,
\begin{align}
    J =  \det\left( \delta_{ij} + \Psi_{i,j} \right)\,.
\end{align}
To solve equation \eqref{eqn:lptmaster} we make the Ansatz, 
\begin{align}
    \Psi_{i,j} &= D_1 \Psi^{(1)}_{i,j} + D_2 \Psi^{(2)}_{i,j} + D_{2\lambda} \Psi^{(2\lambda)}_{i,j} + ...\,, \label{eqn:2lptansatzaniso2}
\end{align}
where $D_1$ and $\Psi^{(1)}$ are given by
\begin{align}
    \myvec{\Psi}^{(1)} &= - \nabla \Phi^{(1)}\,, \\
    \nabla^2 \Phi^{(1)} &= \delta^{(1)} = \delta_{\text{lin}}\,, \,\\
\textrm{and}\,\, \ddot{D}_{1}  +  2 H \dot{D}_{1} &- \frac{3}{2} H^2 \Omega_m D_{1} = 0\,,
\end{align}
where $\nabla^2 \equiv \sum_i \partial_i^2$, 
 which are the same 1LPT solutions as in the standard isotropic frame. $D_2$ and $\Psi^{(2)}$ are given by
\begin{align}
    \myvec{\Psi}^{(2)} &= \hlm{\nabla \Phi^{(2)}}\,,\\
    \nabla^2 \Phi^{(2)} &= \sum_{i > j} \left( \Phi_{,ii}^{(1)} \Phi_{,jj}^{(1)} - \left[\Phi_{,ij}^{(1)}\right]^2  \right) \hlm{+} \left( \sum \Phi_{,ii}^{(1)} \right) \left( \sum \lambda_j \right) \label{eqn:2lptpotential}\,,  \\
\textrm{and}\,\, \ddot{D}_{2}  &+  2 H \dot{D}_{2} - \frac{3}{2} H^2 \Omega_m D_{2} = - \frac{3}{2} H^2 \Omega_m D_1^2\,,
\end{align}
which are identical to the standard $2$LPT solutions except that the 2LPT potential \eqref{eqn:2lptpotential} contains an additional term which is sourced by the large-scale overdensity $\sum \lambda_i$. $D_{2\lambda}$, and $\Psi^{(2\lambda)}_{i,j}$ are also of second order and represent corrections to these solutions which have another time-dependence $D_{2\lambda}$. We expand \eqref{eqn:lptmaster} up to second order with the Ansatz in \eqref{eqn:2lptansatzaniso2}. We use the approximations to $\alpha_i$ and the anisotropic Hubble tensor from \eqref{eqn:approxevol} and  \eqref{eqn:anisohubble} which assume that the large-scale tidal field is caused by large-scale, linear density perturbations and find
\begin{align}
    \ddot{D}_2 \Psi_{i,i}^{(2)} +  2 H \dot{D}_2 \Psi_{i,i}^{(2)} - \frac{3}{2} H^2 \Omega_m \Psi_{i,i}^{(2)} D_2  \nonumber \\
    + \ddot{D}_{2\lambda} \Psi_{i,i}^{(2\lambda)} +  2 H \dot{D}_{2\lambda} \Psi_{i,i}^{(2\lambda)} - \frac{3}{2} H^2 \Omega_m D_{2\lambda} \Psi_{i,i}^{(2\lambda)} \nonumber \\
    = - \frac{3}{2} H^2 \Omega_m D_1^2 \left( \frac{1}{2} \left( \Psi_{k,k}^{(1)} \right)^2 - \frac{1}{2} \Psi_{i,j}^{(1)} \Psi_{j,i}^{(1)}  \right) \nonumber \\
    + 2  \dot{D}_1^2 \sum_i \lambda_i \Psi_{i,i}^{(1)}  \hlm{+ \frac{3}{2} H^2 \Omega_m  D_1^2\left( \sum_i \Psi_{i,i}^{(1)} \right) \sum_j \lambda_j }\,,
\end{align}
which we can simplify with the solutions for $\Psi^{(1)}$ and $\Psi^{(2)}$ to get
\begin{align}
    \ddot{D}_{2\lambda} \Psi_{i,i}^{(2\lambda)} +  2 H \dot{D}_{2\lambda} \Psi_{i,i}^{(2\lambda)} - \frac{3}{2} H^2 \Omega_m D_{2\lambda} \Psi_{i,i}^{(2\lambda)} \nonumber \\
    = 2 \dot{D}_1^2 \sum_i \lambda_i \Psi_{i,i}^{(1)} \,,
\end{align}
where the spatial and time dependent parts can be separated to obtain the differential equations,
\begin{align}
  \myvec{\Psi}^{(2\lambda)} &= \hlm{\nabla \Phi^{(2\lambda)}}\,, \\
  \nabla^2 \Phi^{(2\lambda)} &= \hlm{-} \sum_i \lambda_i \partial_i^2 \Phi^{(1)}
     \label{eqn:d2agrowth}\,, \\
     \textrm{and}\,\, \ddot{D}_{2\lambda} +  2 H \dot{D}_{2\lambda} &- \frac{3}{2} H^2 \Omega_m D_{2\lambda} = 2 \dot{D}_1^2  \,.
     \nonumber
\end{align}
We note that equation \eqref{eqn:d2agrowth} has very similar shape to the differential equation for $D_{2}$. We find that
\begin{align}
    D_{2\lambda} = D_1^2 + D_2
\end{align}
exactly solves equation \eqref{eqn:d2agrowth}. This solution can also be found by doing perturbation theory in the isotropic frame with an additional fixed tidal field source term. This itself gives growth proportional to $D_2$. The coordinate transformation into the anisotropic frame introduces an additional time dependence $\propto D_1^2$ which leads to $D_{2\lambda} = D_2 + D_1^2$. 

\subsection{Useful Approximations} \label{lpt:appendix2lptapprox}
We note that $D_2 \approx - 3/7 D_1^2$ so that
\begin{align}
    D_{2\lambda} \approx - \frac{4}{3} D_2 \approx \frac{4}{7} D_1^2\,.
\end{align}
Since $D_{2\lambda} \approx - \frac{4}{3} D_2$ for an Einstein de Sitter universe, it is also possible to absorb $\Phi_{2\lambda}$ into the potential term that is proportional to $D_2$ if the universe is close to Einstein de-Sitter:
\begin{align}
    \nabla^2 \Phi^{(2*)} &= \sum_{i > j} \left( \Phi_{,ii}^{(1)} \Phi_{,jj}^{(1)} - \left[\Phi_{,ij}^{(1)}\right]^2  \right) \nonumber \\ 
    & \hlm{+} \left( \sum \Phi_{,ii}^{(1)} \right) \left( \sum \lambda_j \right) \hlm{+} \frac{4}{3} \sum \lambda_i \partial_i^2 \Phi^{(1)}\,, \label{eqn:2lptmod}
\end{align}
which requires only minor modifications to a standard 2LPT code for generating initial conditions of an anisotropic universe. 

\begin{figure}
    \centering
    \includegraphics[width=\columnwidth]{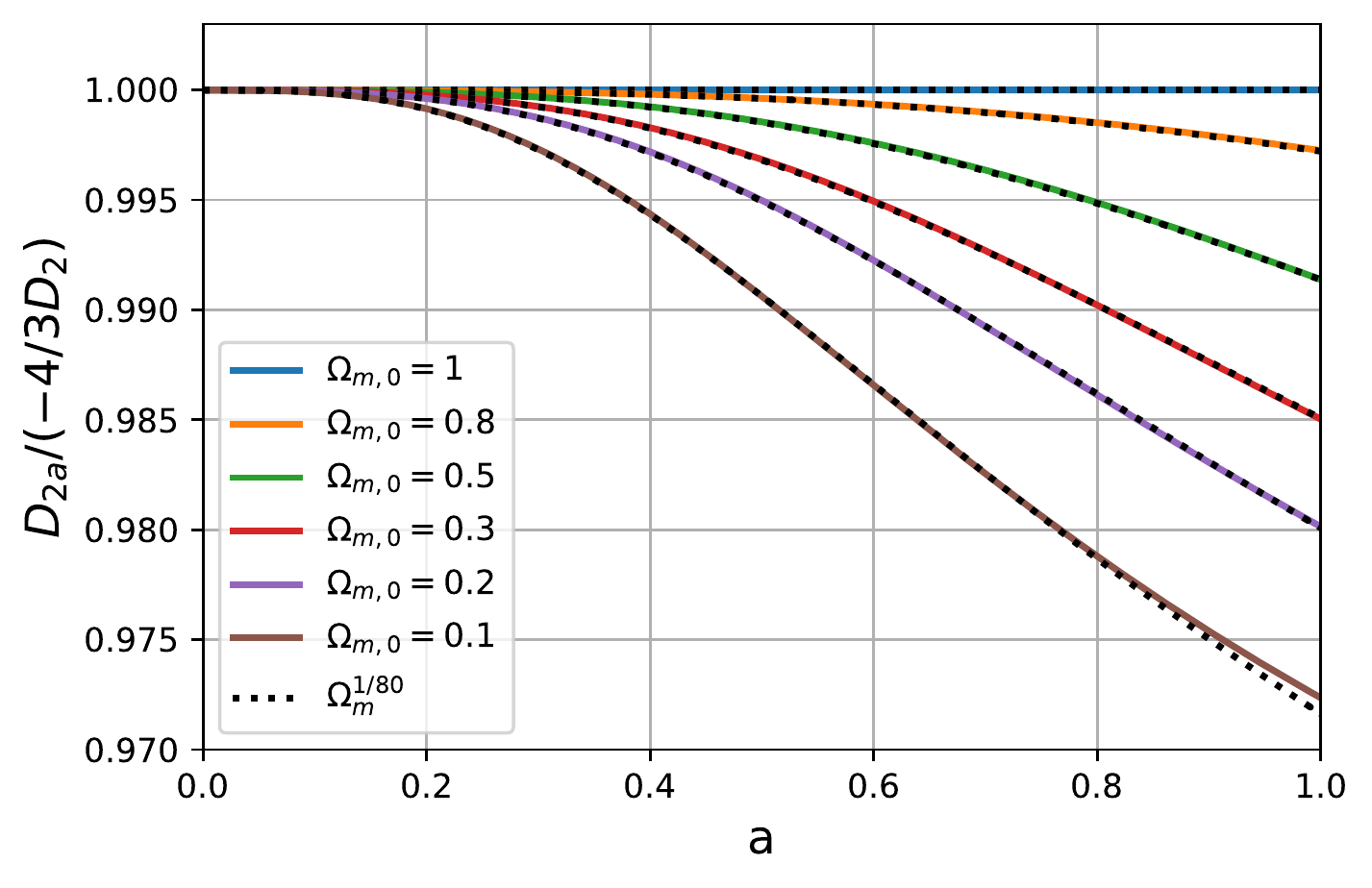}
    \caption{Ratio between the growth factor of the anisotropic response and the Einstein-de-Sitter approximation $D_{2\lambda} \approx -\frac{4}{3} D_2$ for a flat universe, $\Omega_\Lambda = 1 - \Omega_m$, with different values for $\Omega_m$. The difference in response is of order $1.5\%$ at $a=1$ for a sensible cosmology with $\Omega_m \sim 0.3$. We find as a fit $D_{2\lambda} \approx \frac{4}{7} D_1^2 \Omega_m^{1/185} \approx - \frac{4}{3} D_2 \Omega_m^{1/80}$.}
    \label{fig:responce_lcdm}
\end{figure}

As illustrated in Fig.~\ref{fig:responce_lcdm}, the error of this approximation is extremely small and scales roughly as $D_{2\lambda}/(-4/3 D_2) \approx \Omega_m(a)^{1/80}$ (note that we find that $D_{2\lambda}/(4/7 D_1^2) \approx \Omega_m(a)^{1/185}$ describes very well the response in terms of $D_1$). At $z=0$ this is of order a percent, and at early times, $z \sim 100$, it will be completely negligible. We therefore conclude that \eqref{eqn:2lptmod} provides a simple and very accurate modification for the 2LPT initial conditions in the anisotropic frame.


\subsection{Response} \label{app:response}

To validate the 2LPT solution, we estimate the growth-only response for early times. The response can be estimated by looking at the change of the density field for small differences in $\lambda_i$:
\begin{align}
    \frac{\partial \delta}{\partial \lambda_i} &= \hlm{ \frac{\partial J^{-1}} {\partial \lambda_i}} \\
    &= - D_2 \left(\sum_j \phi_{,jj}^{(1)}\right) + D_{2\lambda} \hlm{\phi_{,ii}^{(1)}}  \\
    \hlm{\frac{\partial \delta_k}{\partial \lambda_i}} &\approx \hlm{\frac{\delta_{k}}{D_1} \left(-D_2 + D_{2\lambda} \frac{k_i^2}{k^2} \right) } \\
    \delta_k^2(\myvec{\lambda}) &\approx \left(\delta_k + \sum \lambda_i \frac{\partial \delta_k}{\partial \lambda_i} \right)^2 \\
    &\approx \delta_k^2 + 2 \delta_k \sum \lambda_i \frac{\partial \delta_k}{\partial \lambda_i} \nonumber \\
    &\approx  \hlm{\delta_k^2 \left(1 + 2 \sum_i D_1 \lambda_i \left(-\frac{D_2}{D_1^2} +  \frac{D_1^2 + D_2}{D_1^2} \frac{k_i^2}{k^2} \right) \right)} \nonumber \\
\end{align}
If we split the $\lambda_i$ into an over-density and a trace-free component:
\begin{align}
    \delta^* &\hlm{=} \sum \lambda_i \,,\\
    \lambda^*_i &\hlm{=} \lambda_i - \frac{\sum \lambda_i}{3}\,,
\end{align}
we find 
\begin{align}
    \delta_k^2(\myvec{\lambda}) &= \delta_k^2 \left(1 - 2 \frac{D_2}{D_1^2} D_1 \delta^* + 2 \left(1 + \frac{D_2}{D_1^2}\right) \left( \hlm{\frac{D_1 \delta^*}{3}} + \sum D_1 \lambda_i^* \frac{k_i^2}{k^2} \right) \right) \nonumber \\
    &= \delta_k^2 \left(1 + \left(\hlm{\frac{2}{3}} - \hlm{\frac{4}{3}} \frac{D_2}{D_1^2} \right) D_1 \delta^* + \left(2 + 2 \frac{D_2}{D_1^2}\right) \sum \hlm{D_1 \lambda_i^*} \frac{k_i^2}{k^2} \right) \label{eqn:responseexact} \\
    &\approx \delta_k^2 \left(1 + \hlm{\frac{26}{21}} D_1 \delta^* + \frac{8}{7} \sum D_1  \lambda_i^* \frac{k_i^2}{k^2} \right)\,, \label{eqn:responseeds}
\end{align}
where \eqref{eqn:responseexact}  gives the exact response (also in $\Lambda$CDM) and \eqref{eqn:responseeds} is only exact in an Einstein-de-Sitter universe for which $D_2 \approx -3/7 D_1^2$. The coefficient $8/7$ matches the result known from second-order perturbation theory; in addition, the coefficient of $\hlm{26/21}$ of the term proportional to the long-wavelength density perturbation also matches known perturbation theory results \cite{baldauf/etal:2011,response}.

Based on our numerical results above, we found as a good fit (equivalent to the one in Figure \ref{fig:responce_lcdm}) for the response in a flat $\Lambda$CDM universe:
\begin{align}
    G_{K} \approx \frac{8}{7} \Omega^{1/185} _m(a)\,,
\end{align}
which deviates by less than a percent from $8/7$ at $z=0$ and by much less at earlier times.

\section{Elliptical TreePM}
\label{sec:Elliptical_treepm}
As discussed in section \ref{sec:AnisoTree} we need to calculate the potential of an ellipsoid to infer an expression for the short range force. Here we derive the integral that needs to be solved for an exact evaluation and further show how it can be approximated to a reasonable accuracy through a series expansion.
\subsection{Elliptical potential}\label{sub:EllipticalP}
To find a real-space representation of the long-range potential we have to evaluate $\phi_l = \frac{1}{r} * f$ as a convolution of the Green's function $G  = \frac{1}{r}$ with $f$, which has an ellipsoidal shape with a Gaussian kernel:
\begin{align}
  f(u) &=  \rho_0 \exp[- u^2 / (2 \sigma^2)] \label{eqn:ellipsodialmassapp}\,,\\
  u^2(\myvec{r}) &= \frac{x^2}{a^2} + \frac{y^2}{b^2} + \frac{z^2}{c^2}\,,
\end{align}
where we have written for simplicity $\alpha_1 = a$, $\alpha_2 = b$, $\alpha_3 = c$, and $r_1 = x$, $r_2 = y$ and $r_3 = z$ are isotropic comoving coordinates. Further, $\sigma = \sqrt{2} r_s$. The half-axes of the ellipsoid are thus given by $\alpha_i r_s$.
Therefore, we have to calculate the potential of an ellipsoid which is, following e.g. \cite{1969efe..book.....C, 2008gady.book.....B}, given by
\begin{equation}
\label{eq:LongPot}
\phi_{l} = 2\pi a b c \int_{0}^{\infty}\frac{G(\infty) - G(u(v,x,y,z))}{\sqrt{(a^2 + v)(b^2 + v)(c^2 +v)}} dv\,,
\end{equation}
with 
\[
G(u) = \int_{0}^{u} u' \rho (u')du'\,,
\]
and 
\[
u = \sqrt{\frac{x^2}{a^2+v}+\frac{y^2}{b^2+v}+\frac{z^2}{c^2+v}}\,.
\]
In this specific case, we have for the kernel $\rho (u) = f(u)$.
With that $G(u) , G(\infty)$ can be calculated by
\begin{align}
\label{eq:Gofu}
G(u) &= \rho_{0} \int_{0}^{u} u'\, \exp\left(-\frac{u'^2}{2\sigma^2}\right) du' \\
 &= \rho_{0} \sigma^{2} \left[ 1- \exp\left(-\frac{u^2}{2\sigma^2}\right)\right] \,,\nonumber\\
 G(\infty) &= \rho_{0} \sigma^{2}\,.
\end{align}
Now, inserting \refeq{Gofu} into \refeq{LongPot}, the equation for the potential is found as
\begin{align}
\phi_{l}(\myvec{x}) &= 2\pi a\,b\,c \rho_{0}\sigma^2 \nonumber \\
&\times \int_{0}^{\infty}\frac{\exp\left[-\frac{1}{2\sigma^2} \left(\frac{x^2}{a^2+v} + \frac{y^2}{b^2+v} + \frac{z^2}{c^2+v}\right)\right]}{\left((a^2+v)(b^2+v)(c^2+v)\right)^{1/2}} dv\,.
\label{eq:LongPot2_app}
\end{align}
This is now a function of the three (scaled) axes of the ellipsoid $a,b,c  = \alpha_{0}, \alpha_{1}, \alpha_{2}$ and an elliptical coordinate $v$.
The long range potential \refeq{LongPot2_app} can not be solved analytically.
In this work, we approximate the long range potential by a series expansion around a mean $\bar{\alpha} \equiv (\alpha_{0}\alpha_{1}\alpha_{2})^{1/3}$.
This results in an integral that is independent of the different directions, which means the integral can be calculated once and stored in an interpolation table for different $\bar{\alpha}$.
Only the additional factors from the expansion are dependent on direction and are easy and - most importantly - fast to calculate.
\subsection{Elliptical potential approximation - Series expansion}
\label{sub:EPA}
Following the short derivation of its analytic form, an approximation for the long-range potential is now described.
Since we only expect moderate axis ratios, we expand the integrand in \refeq{LongPot2_app},
\be
\label{eq:integrant}
\zeta = \frac{\exp\left[-\frac{1}{2\sigma^2} \left(\frac{x^2}{a^2+v} + \frac{y^2}{b^2+v} + \frac{z^2}{c^2+v}\right)\right]}{\left((\alpha_{0}^2+v)(\alpha_{1}^2+v)(\alpha_{2}^2+v)\right)^{1/2}}\,,
\ee 
around the spherical case $\alpha_{i} = \bar{\alpha}\, \forall i \in {0,1,2}$.
In the spherical case the integrand (\ref{eq:integrant}) can be simplified to
\be
\label{eq:spherical_integrant}
\zeta|_{\myvec{\alpha} = \bar{\alpha}} = L_{3}\,,
\ee
with 
\be
\label{eq:spherical_integrant_m}
L_{m} = \frac{\exp\left[-\frac{1}{4r_{s}^2} \left(\frac{r^2}{\bar{\alpha}^2+v}\right)\right]}{\left(\bar{\alpha}^2+v\right)^{m/2}}\,,
\ee
where we replaced $\sigma = \sqrt{2} r_{s}$.
The series expansion up to second order is given as
\ba
\zeta \approx\:& L_{3} + \sum_{i} \left.\frac{\partial \zeta}{\partial \alpha_{i}}\right|_{\myvec{\alpha} = \bar{\alpha}} (\alpha_{i} - \bar{\alpha}) \nonumber \\
&+ \frac{1}{2}\sum_{ij}\left.\frac{\partial^2 \zeta}{\partial \alpha_{i}\partial \alpha_{j}}\right|_{\myvec{\alpha} = \bar{\alpha}}(\alpha_{i} - \bar{\alpha})(\alpha_{j} - \bar{\alpha}) + \mathcal{O}(3)\,.
\ea
The first and second order terms are then:
\ba
\label{eq:Firstorder_exp}
\left.\frac{\partial \zeta}{\partial \alpha_{i}}\right|_{\myvec{\alpha} = \bar{\alpha}} &= -L_{5}\bar{\alpha} + L_{7}r_{i}^2 \frac{\bar{\alpha}}{2r_{s}^2} \\ 
              &= L_{5}f_{5,I} + L_{7}f_{7,I}\,,\\
\left.\frac{\partial^2 \zeta}{\partial \alpha_{i}\partial \alpha_{i}}\right|_{\myvec{\alpha} = \bar{\alpha}} & = -L_{5} + L_{7}\left(3\bar{\alpha}^2 + \frac{r_{i}^2}{2r_{s}^2}\right) + \bar{\alpha}^2\left( -L_{9}\frac{3r_{i}^2}{r_{s}^2} + L_{11}\frac{r_{i}^4}{4r_{s}^2}\right)\nonumber \\
&=: f_{5,II}L_{5}+L_{7}f_{7,II,i=j} + L_{9}f_{9,II,i=j} + L_{11}f_{11,II,i=j}\,,\\
\left.\frac{\partial^2 \zeta}{\partial \alpha_{i}\partial \alpha_{j}}\right|_{\myvec{\alpha} = \bar{\alpha}} & \stackrel{i\neq j}{=} \bar{\alpha}^2 \left(L_{7} - L_{9}\frac{r_{i}^2 + r_{j}^2}{2 r_{s}^2} + L_{11}\frac{r_{i}^2 r_{j}^{2}}{4r_{s}^2}\right) \nonumber \\
&= f_{7,II,i\neq j}L_{7} + f_{9,II,i\neq j}L_{9} + f_{11,II,i\neq j}L_{11}\,,
\label{eq:secondorder_Exp}
\ea
where we have given labels to the factors that are associated with different $L_m$: $f_{m,I}$ for the first order terms and $f_{m,II,...}$ for the second order terms. 
Integrating \refeq{spherical_integrant_m} over $v$ we find the solution:
\be
I_{m} = \int_{0}^{\infty}L_{m} \mathrm{d}v = \left(\frac{2r_{s}^2}{r}\right)^{m-2}\left(\Gamma\left[\frac{m-2}{2},0\right] - \Gamma\left[\frac{m-2}{2},\frac{r^{2}}{4\bar{\alpha}^2r_{s}^{2}}\right]\right)\,, \label{eq:Im}
\ee
with the incomplete gamma function $\Gamma$.
Thus the potential Eq. (\ref{eq:LongPot2_app}) can be written to first order as a series in $\Delta\alpha_{i} = \alpha_{i} - \bar{\alpha}$:
\begin{align}
\phi_{l}(\myvec{x}) &\approx 4\pi abc\rho_{0}r_{s}^2 \nonumber \\
&\times\left(I_{3}(r) + \sum_{i} \left( -I_{5}(r) \bar{\alpha} + I_{7} r_{i}^2 \frac{\bar{\alpha}}{2r_{s}^2}\right)\overbrace{(\alpha_{i} - \bar{\alpha})}^{\equiv\Delta \alpha_{i}} + ...\right)\,.
\label{eq:phi_long_app}
\end{align}
or we write as an abbreviation for the potential to higher orders
\begin{align}
    \phi_{l}(\myvec{x}) &\approx 4\pi abc\rho_{0}r_{s}^2 \sum f_m I_m
\end{align}
where the sum goes over all required terms for a given order. The $f_m$ (as defined above up to second order) have general (but analytic) coordinate dependencies. The $I_m$ depend only on the radial coordinate.

\subsection{Force equations}
The next step is to derive the equation for the force. We have to take the derivative with respect to anisotropic comoving coordinates which is related to the gradient in isotropic coordinates in a simple manner:
\begin{align}
    F_k^{\text{long}} = -\frac{\partial \phi_l}{\partial x_k} &= - \hlm{\alpha_k} \frac{\partial \phi_l}{\partial r_k} =: - \hlm{\alpha_k} \partial_{r_k} \phi
\end{align}
The gradient in isotropic coordinates is given by
\begin{equation}
\label{eq:force_main}
\partial_{r_k} \phi_{l} = 4\pi abc\rho_{0}r_{s}^2\left(\frac{r_k}{r} \sum f_{m} \partial_{r} I_{m}(r) + \sum I_{m} \partial_{r_k} f_{m}\right)\,,
\end{equation}
with the factors $f_{m}$ from \refeqs{Firstorder_exp}{secondorder_Exp} and the derivative of $I_{m}$,
\begin{align}
I'_{m} := \partial_r I_{m} &= \frac{1}{4 r}\left(\frac{r_{s}}{r}\right)^{m} \left(8 \bar{\alpha}^2 \exp\left(-\frac{r^2}{4\bar{\alpha}^2 r_{s}^2}\right) \left(\frac{r^2}{\bar{\alpha}^2 r_{s}^2}\right)^{m/2}\right. \nonumber \\
&-\left. \frac{2^m (m-2)r^2 \left( \Gamma\left[\frac{m-2}{2}, 0\right] - \Gamma\left[\frac{m-2}{2}, \frac{r^2}{4\bar{\alpha}^2 r_{s}^2}\right]\right)}{r_{s}^2}\right)\,.
\end{align}
The derivatives for $f_{m}$ with respect to $r_{k}$ are:
\begin{align}
\partial_{r_{k}} f_{3} &= 0\,,\\
\partial_{r_{k}} f_{5} &= 0\,,\\
\partial_{r_{k}} f_{7, I} &= \delta_{ik} \frac{\bar{\alpha}}{r_{s}^2} \Delta\alpha_{i} r_{i}\,,\\
\partial_{r_{k}} f_{7,II, i=j} &= \delta_{ik} \frac{r_{i}}{r_{s}^2}\,,\\
\partial_{r_{k}} f_{7, II i\neq j} &= 0\,,\\
\partial_{r_{k}} f_{9, II, i=j} &= -\delta_{ik} \frac{6r_{i}}{r_{s}^2} \bar{\alpha}^2\,,\\
\partial_{r_{k}} f_{9, II, i\neq j} &= - \frac{\bar{\alpha}^2}{r_{s}^2} \left( \delta_{ik} r_i + \delta_{jk} r_k \right)  \,,\\
\partial_{r_{k}} f_{11, II, i=j} &= \delta_{ik} \frac{r_{i}^3}{r_{s}^4} \bar{\alpha}^2\,,\\
\partial_{r_{k}} f_{11, II, i\neq j} &= \frac{\bar{\alpha}^2}{2r_{s}^4} \left( \delta_{ik} r_{i} r_{j}^2 + \delta_{jk} r_{i}^2 r_{j} \right)\,,
\end{align}
Therefore  the potential gradient up to first order is given by
\begin{align}
\frac{\partial\phi_l}{\partial r_{k}} &= 4\pi abc\rho_{0}r_{s}^2\left[\frac{r_{k}}{r}\left(I'_{3} - \bar{\alpha} I'_{5} \sum_{i}\Delta\alpha_{i} + \frac{\bar{\alpha}}{2r_{s}^2} I'_{7} \left(\sum_{i}(\Delta\alpha_{i} r_{i}^2)\right)\right)\right. \nonumber \\
&\left.+\frac{\bar{\alpha}}{r_{s}^2} I_{7} \Delta\alpha_{k}r_{k}\right]\,,
\end{align}
with the difference from the mean, $\Delta \alpha_{k} = \alpha_{k} - \bar{\alpha}$. The second-order terms are
\begin{align}
\frac{r_{k}}{2 r} \sum_{i}\sum_{j} &\left(\Delta\alpha_{i}\Delta\alpha_{j} \left(f_{5} I'_{5} + f_{7} I'_{7} + f_{9} I'_{9} + f_{11} I'_{11}\right) \right) \nonumber \\
+ \frac{1}{2} \sum_{i}\sum_{j} &\left(\Delta\alpha_{i}\Delta\alpha_{j} (\partial_{r_k} f_{5}) I_{5} + (\partial_{r_k}f_{7}) I_{7} + (\partial_{r_k} f_{9}) I_{9} \right. \nonumber \\ & \left. + (\partial_{r_k} f_{11}) I_{11}\right)\,.
\end{align}
These equations can be further simplified in order to allow better implementation into the code, and give very good approximations for the force in the anisotropic case for the axis ratios we use.
Knowing the long-range force, the short-range force can easily be calculated  due to the force split in \gadget.\\

\hlcom{Rewritten this paragraph:} As elaborated in this section, the force-split is chosen to be spherical in the anisotropic comoving frame. Therefore the force transition scale and the truncation scale of the tree force are spherical in the anisotropic frame and ellipsoidal in the isotropic comoving frame. However, there is an additional numerical choice to be made for the softening. We choose to make the softening spherical in the isotropic comoving frame. This is the simpler choice, since it ensures that the potential of a particle becomes identical to that of a point mass beyond the softening radius.


\subsection{Long-range potential}
\label{sec:AppendixPot}
Using the expansion above, we find the following expression for the long-range potential,
\begin{align}
\phi_l &\approx I_{3} - I_{5} \alpha(\Delta\alpha_{1} + \Delta\alpha_{2} + \Delta\alpha_{3}) + I_{7} \frac{\alpha}{2 r_{s}^2}(\Delta\alpha_{1}r_{1}^2 + \Delta\alpha_{2}r_{2}^2 + \Delta\alpha_{3}r_{3}^2) \nonumber\\
&+ \frac{\Delta\alpha_{1}^2}{2}\left(-I_{5} + I_{7} \left(3\alpha^2 + \frac{r_{1}^2}{2r_{s}^2}\right) - I_{9} \frac{3 r_{1}^2}{r_{s}^2} \alpha^2 + I_{11} \frac{r_{1}^4}{4r_{s}^4}\alpha^2\right) \nonumber\\
&+ \frac{\Delta\alpha_{2}^2}{2}\left(-I_{5} + I_{7} \left(3\alpha^2 + \frac{r_{2}^2}{2r_{s}^2}\right) - I_{9} \frac{3 r_{2}^2}{r_{s}^2} \alpha^2 + I_{11} \frac{r_{2}^4}{4r_{s}^4}\alpha^2\right) \nonumber\\
&+ \frac{\Delta\alpha_{3}^2}{2}\left(-I_{5} + I_{7} \left(3\alpha^2 + \frac{r_{3}^2}{2r_{s}^2}\right) - I_{9} \frac{3 r_{3}^2}{r_{s}^2} \alpha^2 + I_{11} \frac{r_{3}^4}{4r_{s}^4}\alpha^2\right) \nonumber\\
&+ \frac{\Delta\alpha_{1}\Delta\alpha_{2}}{2}\left(I_{7}\alpha^2 - I_{9}\alpha^2 \left(\frac{r_{1}^2 + r_{2}^2}{2r_{s}^2}\right) + I_{11}\alpha^2\frac{r_{1}^2r_{2}^2}{4r_{s}^4} \right)\nonumber \\
&+ \frac{\Delta\alpha_{1}\Delta\alpha_{3}}{2}\left(I_{7}\alpha^2 - I_{9}\alpha^2 \left(\frac{r_{1}^2 + r_{3}^2}{2r_{s}^2}\right) + I_{11}\alpha^2\frac{r_{1}^2r_{3}^2}{4r_{s}^4} \right)\nonumber \\
&+ \frac{\Delta\alpha_{2}\Delta\alpha_{3}}{2}\left(I_{7}\alpha^2 - I_{9}\alpha^2 \left(\frac{r_{2}^2 + r_{3}^2}{2r_{s}^2}\right) + I_{11}\alpha^2\frac{r_{2}^2r_{3}^2}{4r_{s}^4} \right)\nonumber \\
&+ \frac{\Delta\alpha_{3}\Delta\alpha_{2}}{2}\left(I_{7}\alpha^2 - I_{9}\alpha^2 \left(\frac{r_{3}^2 + r_{2}^2}{2r_{s}^2}\right) + I_{11}\alpha^2\frac{r_{3}^2r_{2}^2}{4r_{s}^4} \right)\nonumber \\
&+ \frac{\Delta\alpha_{2}\Delta\alpha_{1}}{2}\left(I_{7}\alpha^2 - I_{9}\alpha^2 \left(\frac{r_{2}^2 + r_{1}^2}{2r_{s}^2}\right) + I_{11}\alpha^2\frac{r_{2}^2r_{1}^2}{4r_{s}^4} \right)\nonumber \\
&+ \frac{\Delta\alpha_{3}\Delta\alpha_{1}}{2}\left(I_{7}\alpha^2 - I_{9}\alpha^2 \left(\frac{r_{3}^2 + r_{1}^2}{2r_{s}^2}\right) + I_{11}\alpha^2\frac{r_{3}^2r_{1}^2}{4r_{s}^4} \right)\,,
\end{align}
which can be simplified by merging terms.

\subsection{Approximation Error} \label{app:approximation_error}
\begin{figure}
    \centering
    \includegraphics[width=\columnwidth]{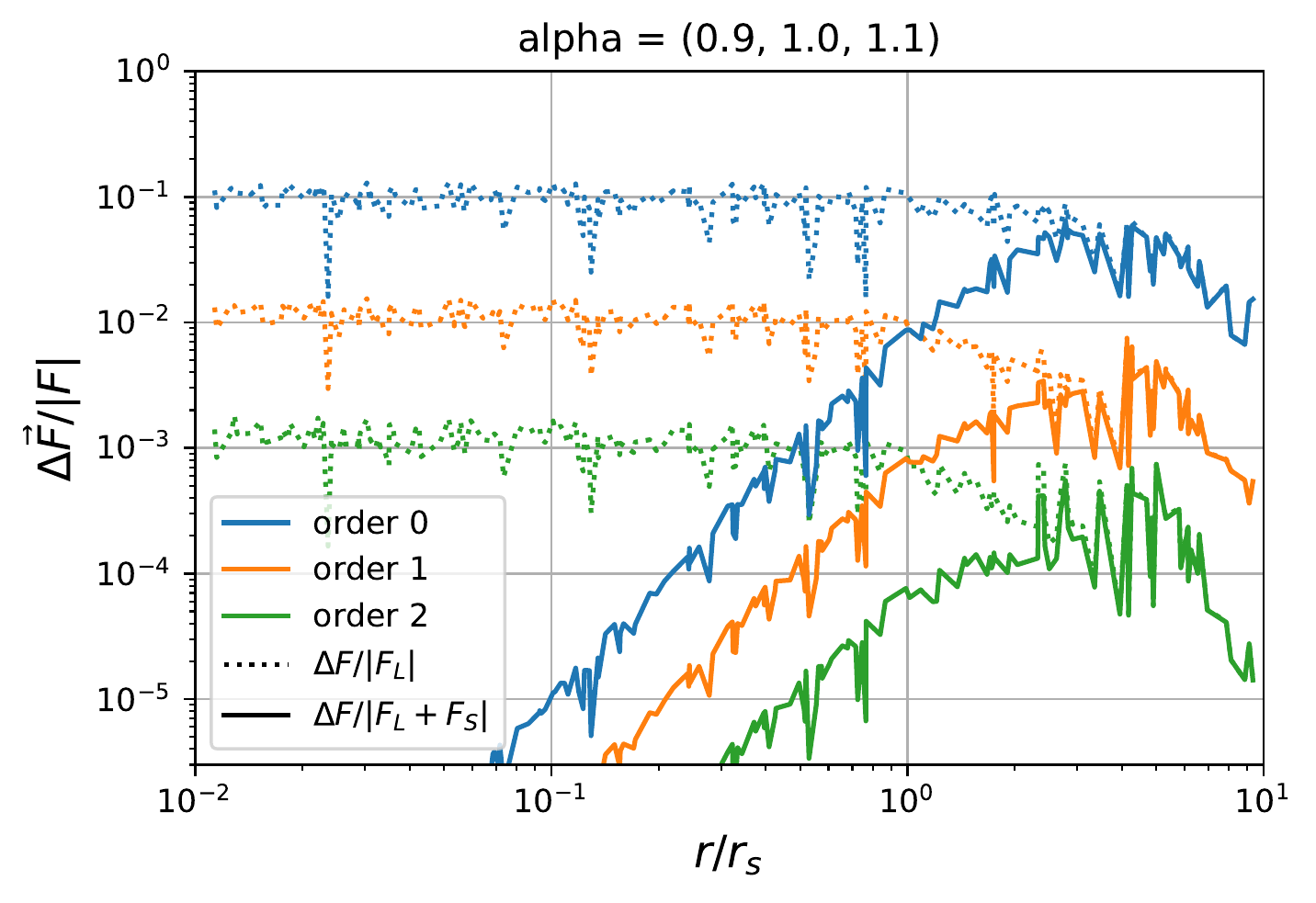}
    \caption{\hlcom{Added solid lines to the figure, adapted caption.} Relative error of the force approximation as a function of radius for the axes $\myvec{\alpha} = (0.9, 1.0, 1.1)^T$. The dotted lines show the error between approximated and true force of the ellipsoid (= $F_L$) and the solid lines show the error relative to the full force estimate $F_{\rm{s}} + F_{\rm{L}}$. This choice of $\myvec{\alpha}$ has already slightly larger axis ratios than our typical simulations, but it still has relative errors smaller than $10^{-3}$ at all radii.}
    \label{fig:forceerrorvsr}
\end{figure}

Here we evaluate the error in our approximate real-space representation of the long-range force. We evaluate the exact force $\myvec{F}$ through numerical integration (and differentiation) of equation \eqref{eq:LongPot2_app} and compare it to our Taylor-approximated version from equation \eqref{eq:force_main} for different approximation orders.  In Figure \ref{fig:forceerrorvsr} we show the relative error for the case of $\myvec{\alpha} = (0.9, 1, 1.1)^T$ for different orders of the expansion. This case has already slightly larger axis ratios than we typically have in our simulations. The force error is evaluated at random directions at a distance $r$.\hlcom{Modified the following sentences: }The dotted lines show the error relative to the true force of the ellipsoid and help to understand the nature of the approximation error. The solid lines show the error relative to the total force (= short-range + long-range) and therefore give an estimate of the relevance of the force-error for the simulations.  The errors seems to be relatively independent of the angle. For the second-order expansion the error is well below $10^{-3}$ at most radii. We can clearly see that the expansion up to second order pays off.

\begin{figure}
    \centering
    \includegraphics[width=\columnwidth]{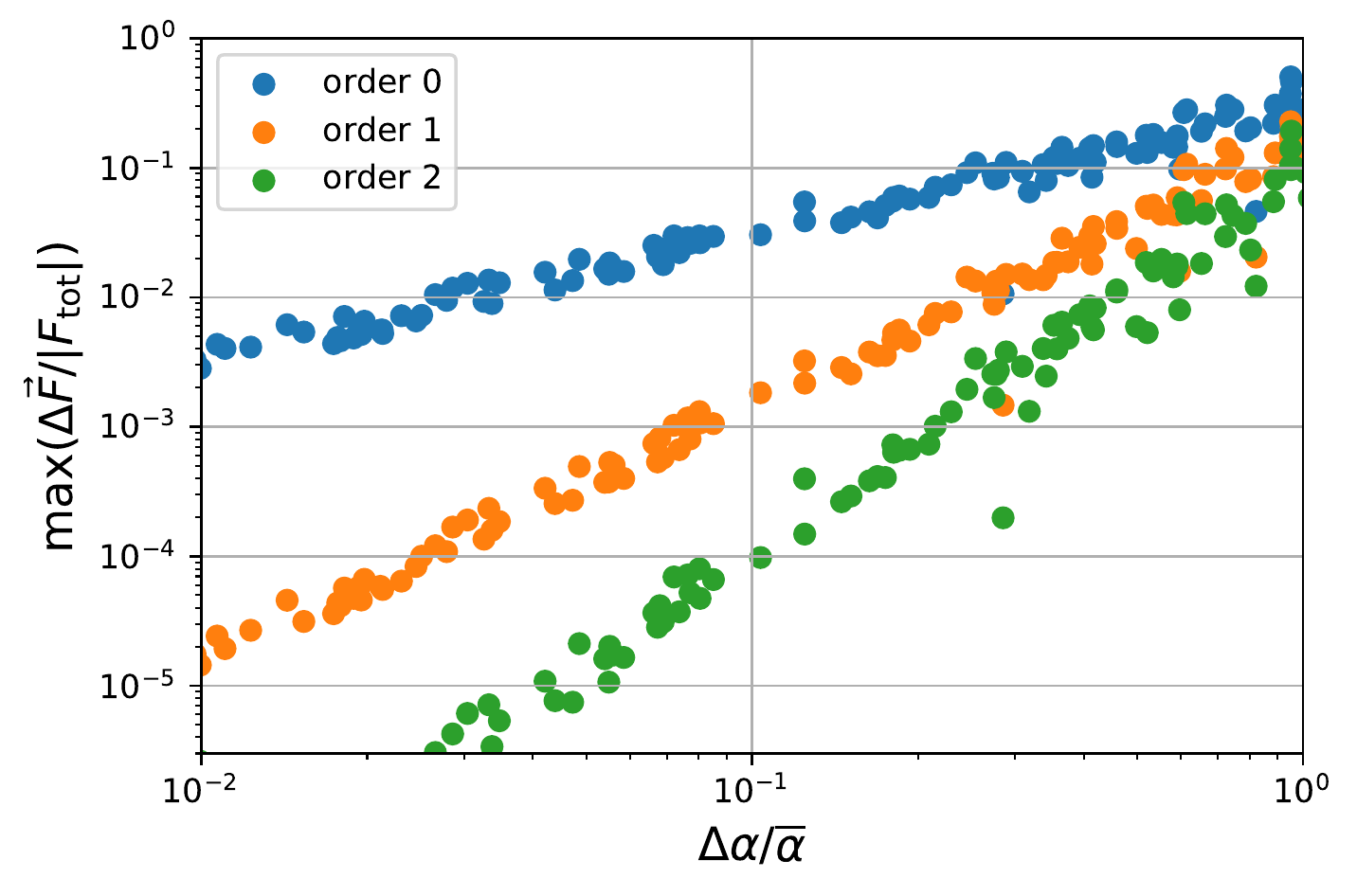}
    \caption{\hlcom{Changed this to $\Delta F/ \Delta F_{\rm{tot}}$. }Maximal relative error $\Delta F/ \Delta F_{\rm{tot}}$ of the force cut on the interval $[0.01 r_s, 10 r_s]$ for different axis ratios. The x-axis indicates the difference between the largest and smallest axis (normalized to the mean axis) and approximately $\Delta \alpha \approx \lambda$. For example, in our simulations with $\lambda \sim 10^{-1}$ we expect force errors well below $10^{-3}$. However, simulations with axis ratios of order unity could not be simulated well with the current approximation.}
    \label{fig:forceerrorvsalpha}
\end{figure}

Further we test how the accuracy of the force approximation behaves as a function of the axis ratios. It is expected that the approximation gets worse if the differences between the axes gets large and we want to check quantitatively in which regime our approximations are good enough. We sample a large number of realizations for the axes $\myvec{\alpha}$ and for each realization we determine the maximum force error on the interval $r \in [0.01 r_s, 10 r_s]$ (the maximum of \hlt{the solid lines as} seen in Figure \ref{fig:forceerrorvsr}). We then plot the maximal relative force error versus the (normalized) difference between the largest and smallest axis as can be seen in Figure \ref{fig:forceerrorvsalpha}. For example the case from Figure \ref{fig:forceerrorvsr} lands at $\Delta \alpha / \overline{\alpha} = 0.2$.

We find that the second order expansion is very accurate for cases considered in this paper with $\lambda \sim \Delta \alpha \sim 0.1$. In future studies we could still consider the approximation reasonably accurate up to $\Delta \alpha \approx 0.4$.\footnote{We remind the reader here that Figure \ref{fig:forceerrorvsalpha} shows the maximum approximation error. Typical errors are smaller.}. For simulations with larger axis ratios than that, more accurate approximations would be needed.

\section{Convergence of the response measurement}
\label{app:response_measurement}

\begin{figure*}
    \centering
    \includegraphics[width=0.8\textwidth]{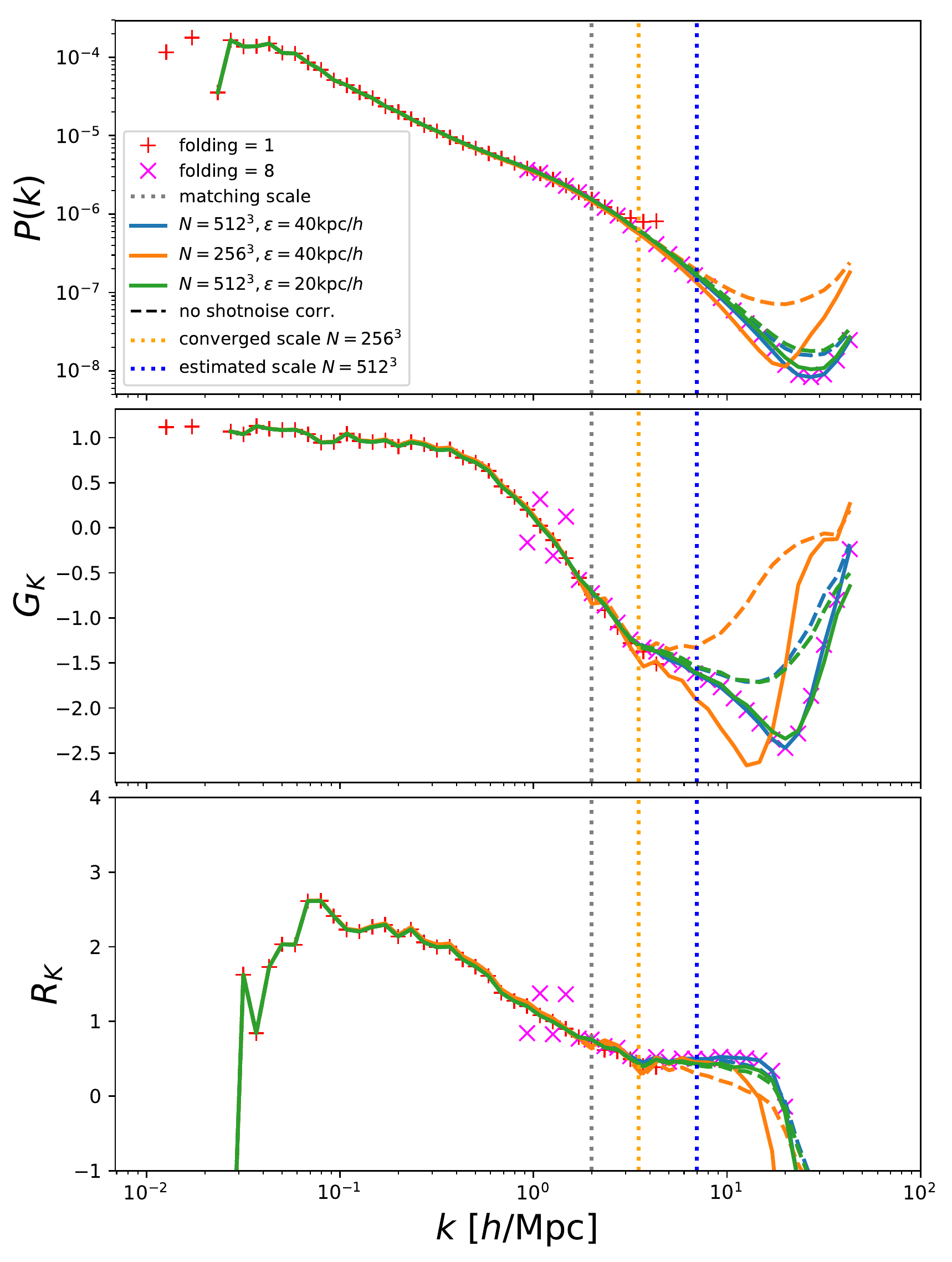}
    \caption{\hlcom{Changed convergence indicating lines. Also adapted caption.}The power spectrum (top), the $G_{K}$-response (center) and the $R_{K}$-response (bottom), showing how they are influenced by numerical details. The $N=256^3$ case is in very good agreement with the higher resolution case up to a scale of $k = \SI{3.5}{\h\per\mega\parsec}$. We estimate that the $N=512^3$ simulations should be reliable up to roughly $k = \SI{7}{\h\per\mega\parsec}$. Please refer to the sections \ref{sec:folding}-\ref{sec:softeningandparticlenumber} for  details about the meaning of individual lines and markers.}
    \label{fig:response_measurement}
\end{figure*}

Here we discuss how to infer the response on small scales by employing a folding technique. Folding techniques have been benchmarked previously on power spectra, but never on the response measurement. Further, we test the impact of a shot-noise correction and of numerical simulation parameters, such as the softening and the particle number. All these tests are presented in Figure \ref{fig:response_measurement}.

\subsection{Folding}
\label{sec:folding}

For all lines presented in Figure \ref{fig:response_measurement} we use a cloud-in-cell (cic) assignment of all simulation particles onto a periodic mesh with a subsequent deconvolution of the cic-kernel in Fourier space to obtain the Fourier-representation of the density field.

To measure the response up to small scales it is necessary to have a Fourier representation which extends up to those scales. For our boxes with $\SI{500}{\per\h\mega\parsec}$ side-length, it is hard to achieve this with a single mesh. We therefore combine two meshes using a folding approach. The first mesh has $N=512$ cells and a side-length of $\SI{500}{\per\h\mega\parsec}$ - giving an accurate measurement of quantities in Fourier space up to $k \sim \SI{2}{\h\per\mega\parsec}$. For the second mesh the density field is folded periodically by a factor 8 onto a grid of side-length  $\SI{62.5}{\per\h\mega\parsec}$. Results for the two meshes  are indicated by crosses and pluses in Figure \ref{fig:response_measurement} for the default case with $N= 512^3$ and $\epsilon =  \SI{40}{\per\h\kilo\parsec}$. The solid and dashed lines show combinations of the two foldings, switching from the unfolded to the folded representation at a matching scale of $k = \SI{2}{\h\per\mega\parsec}$ where the two meshes give reasonably similar results.

\subsection{Shot-Noise correction}
For a uniform Poisson distribution the power spectrum is given by a constant,
\begin{align}
    \langle \delta_k^2 \rangle &= \frac{1}{N}\,,
\end{align}
where $N$ is the number of particles that have been used to infer the density field. It is common to subtract this ``shot-noise" contribution to obtain a more accurate Fourier representation of the underlying density field. We therefore subtract $1/N$ from $\delta_k^2$ before the response-binning to obtain the solid lines in Figure \ref{fig:response_measurement}. We also present the power spectra and response measurements without shot-noise corrections as dashed lines. The shot-noise correction becomes substantial at small scales which are, in any case, subject to other uncertainties, and as a result, it is not entirely necessary.

\subsection{Softening and Particle Number}
\label{sec:softeningandparticlenumber}
The blue lines in Figure \ref{fig:response_measurement} represent our fiducial case of $N=512^3$ particles and a softening of $\epsilon = \SI{40}{\per\h\kilo\parsec}$. The orange line represents a reduction of the particle number by a factor of 8. \hlcom{Rewritten following sentences:} It first deviates from the $N=512^3$ case by more than $0.1$ in $G_k$ at a scale of $k = \SI{3.5}{\h\per\mega\parsec}$. We use this to estimate that the $N=512^3$ simulations are reliable at least up to a wavenumber twice as large, $k = \SI{7}{\h\per\mega\parsec}$, since the Nyquist frequency is also larger by a factor $2$.  

The green line shows a variation of the softening by a factor 2 and we find that this leads to no change on scales larger than $k = \SI{10}{\h\per\mega\parsec}$ and only to minor differences at higher wavenumber.

We conclude that \hlt{we consider} our response measurements reliable up to a scale of $k = \hlt{\SI{7}{\h\per\mega\parsec}}$. This scale is indicated in the plots in section \ref{sec:Results} of the main text.

\bsp	
\label{lastpage}
\end{document}